\documentclass[12pt]{iopart}
\usepackage{graphicx}
\usepackage{graphics}
\usepackage{appendix}
\usepackage{multirow}
\usepackage{hhline}
\usepackage[normalem]{ulem}
\usepackage[usenames, dvipsnames]{color}

\usepackage{hyperref}
\hypersetup{
    colorlinks=true,
    linkcolor=blue,
    filecolor=blue,      
    urlcolor=blue,
    citecolor=blue
}

\begin{document}

\title[Synchrotron spectra in Alcator C-Mod]{Measurements of runaway electron synchrotron spectra at high magnetic fields in Alcator C-Mod}

\author{R.A. Tinguely$^1$\footnote{Author to whom correspondence should be addressed: rating@mit.edu}, R.S. Granetz$^1$, M. Hoppe$^2$, and O. Embr\'{e}us$^2$}

\address{$^1$ Plasma Science and Fusion Center, Massachusetts Institute of Technology, Cambridge, MA, USA \\
$^2$ Department of Physics, Chalmers University of Technology, G\"{o}teborg, Sweden}

\begin{abstract}
In the Alcator C-Mod tokamak, runaway electron (RE) experiments have been performed during low density, flattop plasma discharges at three magnetic fields: 2.7, 5.4, and 7.8 T, the last being the highest field to-date at which REs have been generated and measured in a tokamak.  Time-evolving synchrotron radiation spectra were measured in the visible wavelength range ($\lambda \approx$ 300-1000 nm) by two absolutely-calibrated spectrometers viewing co- and counter-plasma current directions. In this paper, a test particle model is implemented to predict momentum-space and density evolutions of REs on the magnetic axis and $q$~=~1, 3/2, and 2 surfaces. Drift orbits and subsequent loss of confinement are also incorporated into the evolution. These spatiotemporal results are input into the new synthetic diagnostic SOFT [M. Hoppe, \textit{et al.}, Nucl. Fusion \textbf{58}(2), 026032 (2018)] which reproduces experimentally-measured spectra. For these discharges, it is inferred that synchrotron radiation dominates collisional friction as a power loss mechanism and that RE energies decrease as magnetic field is increased. Additionally, the threshold electric field for RE generation, as determined by hard X-ray and photo-neutron measurements, is compared to current theoretical predictions.
\end{abstract}



\section{Introduction}

In a plasma, the interaction of particles via the Coulomb force leads to the interesting characteristic of their collision frequency decreasing with their speed as $1/v^3$. Thus, subject to an electric field $E$, particles of mass $m$ and charge $e$ with velocities $v \geq v_c = (e^3 n_e \ln \Lambda/4\pi\epsilon_0^2 m E)^{-1/2}$ \cite{dreicer1959,dreicer1960} will be continuously accelerated in a plasma of electron density $n_e$ as the driving electric force dominates collisional friction. Due to their light mass, electrons are most susceptible to this phenomenon and are then called ``runaway" electrons (REs). 

In a tokamak, there are four typical scenarios in which REs can be generated. During plasma start-up, the ohmic electric field that ramps the plasma current $I_P$ can produce REs; oftentimes, these are soon quenched due to the subsequent increase in $n_e$ and decrease in loop voltage. However, if $n_e$ remains sufficiently low, REs can persist or form during the flattop current. Additionally, preferential heating of electrons through lower hybrid (LH) current drive or electron cyclotron heating will increase the population of non-thermal electrons which can run away. Yet it is RE beams resulting from disruptions that are usually of most concern. During a disruption, the fast decay of $I_P$ induces a strong electric field that can accelerate REs to energies of tens of MeV. In addition, post-disruption REs have been observed to carry a significant fraction of the the total pre-disruption plasma current: as much as 50\% in ASDEX-Upgrade and 66\% in COMPASS \cite{plyusnin2018}, 60\% in JET \cite{plyusnin2006}, and 80\% in FTU \cite{martin-solis2006}. Loss of confinement of these REs to the first wall can cause serious damage to plasma-facing components (PFCs) and threaten to halt operation of future fusion devices like ITER.

\begin{figure*}[h!]
	\centering
	\includegraphics[width=\linewidth]{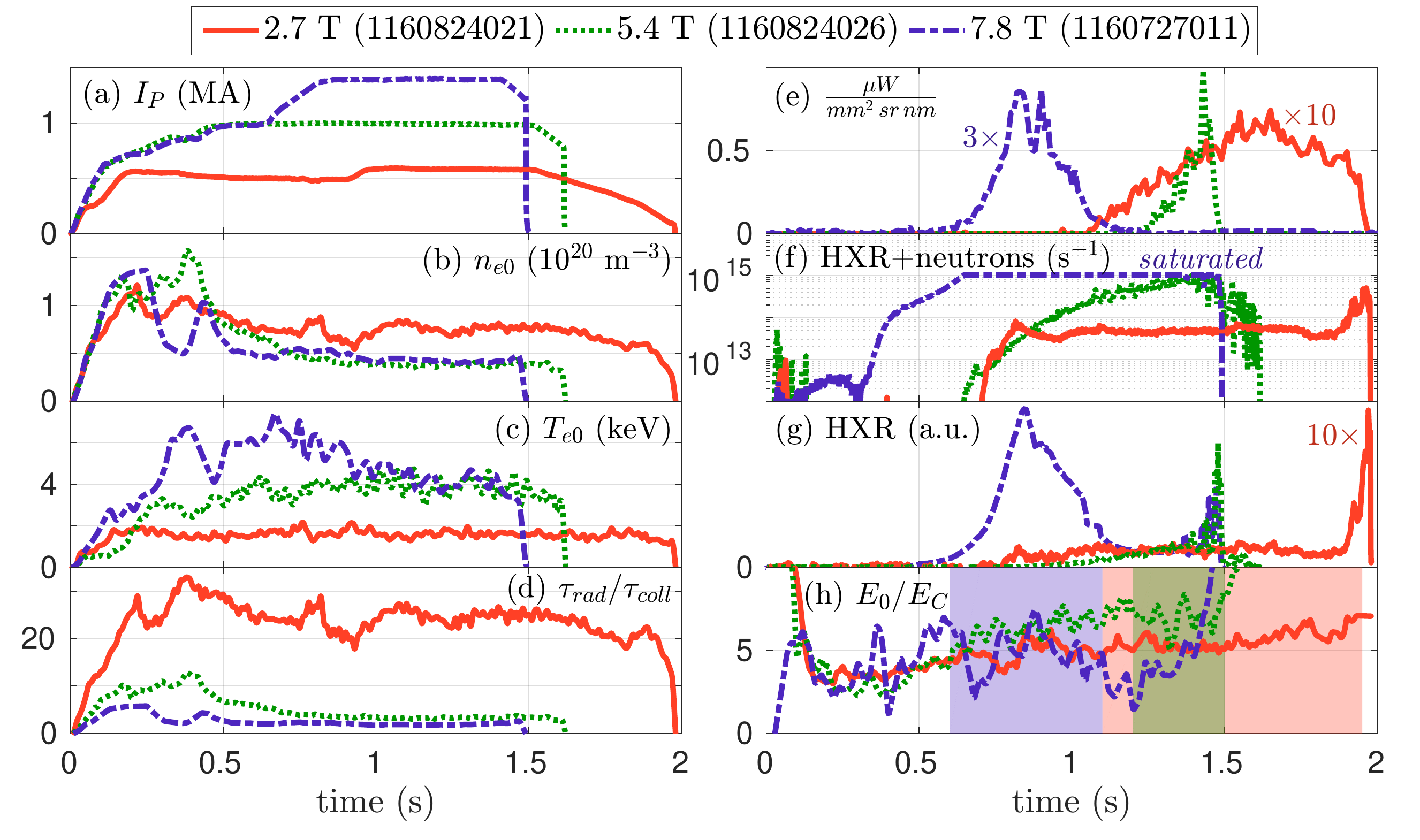}
	\caption{Parameters for three plasma discharges with magnetic field strengths on-axis of 2.7 T (solid), 5.4 T (dotted), and 7.8 T (dot-dashed): (a) plasma current, (b) central electron density, (c) central electron temperature, (d) ratio of the synchrotron radiation timescale to the collisional timescale, (e) measured synchrotron brightness at wavelength $\lambda = 850$ nm, (f) signal resulting from both HXR and photo-neutrons on a logarithmic scale, (g) HXR signal, and (h) ratio of the electric field on-axis to the critical electric field \cite{connor1975}. The shaded regions in (h) highlight the time windows in (e) during which synchrotron radiation is observed. Note that in (e)-(h), some data has been scaled by the factors given.}
	\label{fig:plasmaParameters}
\end{figure*}

\begin{figure}[h!]
	\centering
		\includegraphics[width=0.5\linewidth]{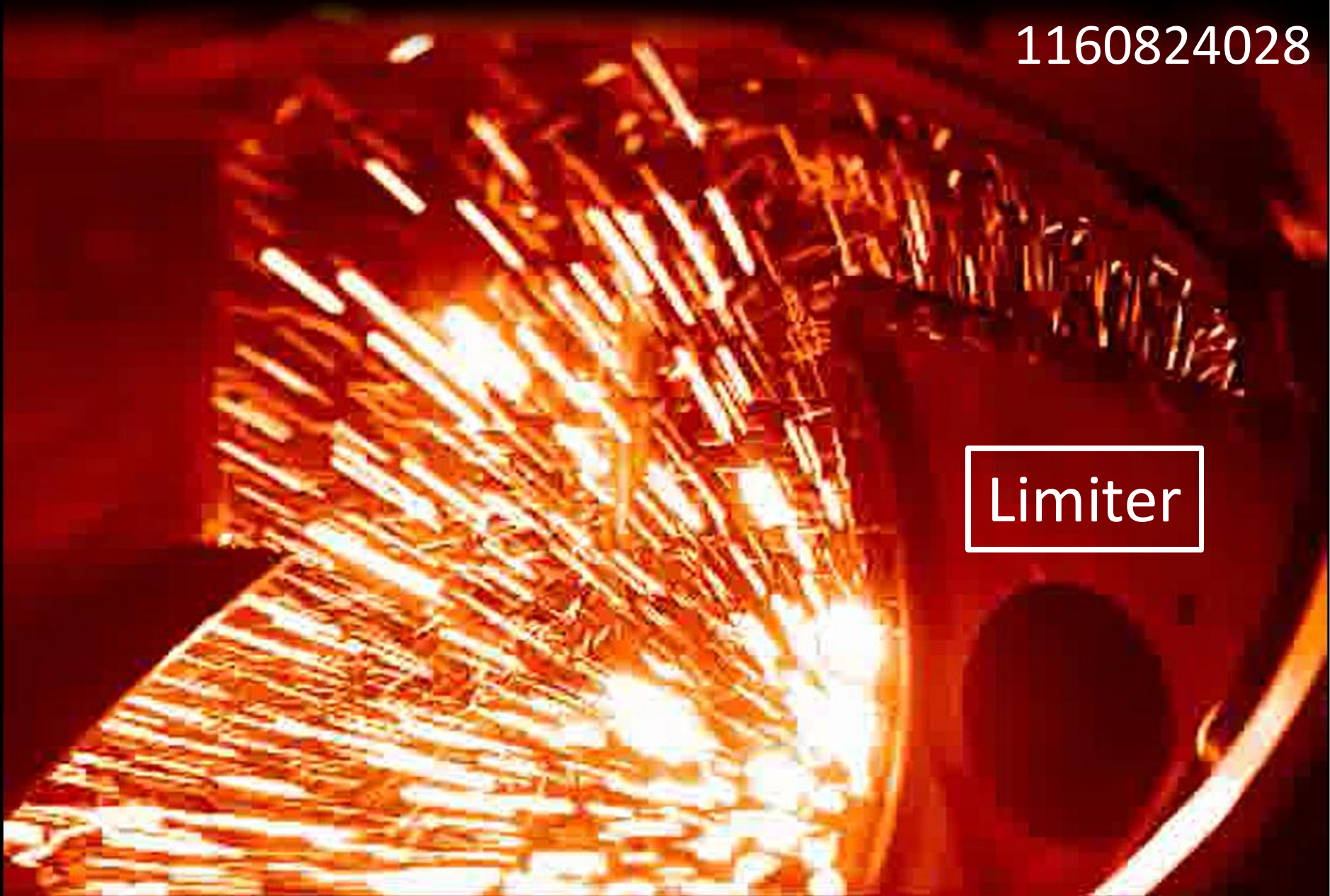}
	\caption{A false-colored visible camera image shows a shower of sparks resulting from REs impacting a limiter during the plasma current ramp-down at the end of an Alcator C-Mod discharge.}
	\label{fig:limiter}
\end{figure}

In the high field, compact Alcator C-Mod tokamak ($B_0 \sim$ 2-8 T, $I_P \sim$ 1-2 MA, $\bar{n}_e \sim 10^{20}$ m$^{-3}$, $R_0$ = 68 cm, $a$ = 22 cm), REs are not observed after disruptions of diverted plasmas, likely due to the fast stochasticization of magnetic flux surfaces \cite{marmar2009,izzo2011}. However, they can be generated during flattop discharges of sufficiently low plasma density ($\bar{n}_e \sim$ (1-5)$\times 10^{19}$ m$^{-3}$), sometimes with the aid of LH seeding or short ($\sim$100 ms) ramps in $I_P$ to increase the loop voltage. This paper explores the effect of varying toroidal magnetic field strength $B$ on the evolution of flattop REs in C-Mod. Figure~\ref{fig:plasmaParameters} shows plasma parameters for three specific discharges of interest during which RE synchrotron radiation spectra were measured. Note also that even flattop REs can still damage PFCs after losing confinement; in figure~\ref{fig:limiter}, REs strike a molybdenum-tiled limiter in C-Mod, producing a spray of molten metal.

In a magnetized plasma, the gyromotion of charged particles about magnetic field lines produces cyclotron radiation. The relativistic extension of this cyclotron emission is called synchrotron radiation and is highly peaked in the direction of particle motion. Due to C-Mod's high magnetic field, a non-negligible fraction ($\sim$5-10\%) of the synchrotron radiation spectrum is observed in the visible wavelength range (see figure~\ref{fig:spectraPankratov}), making it a good indicator of REs as it is seen from only one direction in the tokamak (counter-$I_P$). In addition, synchrotron emission can provide valuable information about the energy and number of REs, as well as their distribution throughout the plasma. Understanding RE dynamics in both momentum and real space allows the validation of theory, simulations, and synthetic diagnostics as well as better prediction, detection, and mitigation of REs in future fusion devices. 

In this paper, we explore synchrotron radiation specifically as a power loss mechanism by generating REs and measuring their synchrotron spectra at three magnetic fields: $B_0$ = 2.7, 5.4, and 7.8 T. To the authors' knowledge, 7.8 T is the highest magnetic field at which REs have been generated and observed in a tokamak. A test particle model of energy \cite{martin-solis1998} and density \cite{connor1975,rosenbluth1997} evolution is coupled with the new synthetic diagnostic SOFT \cite{hoppe2017} to produce synthetic spectra which can be directly compared with experiment.  For the first time, spatial effects are included in the RE evolution and resulting spectra. Good agreement is seen between experimental and predicted spectra, and it is shown that higher magnetic fields are consistent with greater synchrotron power loss and therefore lower RE energies.

The organization of the rest of the paper is as follows: Section~\ref{sec:threshold} investigates the threshold electric field required to generate REs at high $B$ in C-Mod. In section~\ref{sec:synchrotron}, a basic review of synchrotron radiation theory is presented. Experimental efforts on C-Mod are detailed in section~\ref{sec:setup}, and the approach for data analysis is given in section~\ref{sec:approach}. Section~\ref{sec:analysis} expands upon the specific analyses for three discharges, each with a different magnetic field, and the results are discussed in section~\ref{sec:discussion}. Finally, a summary is given in section~\ref{sec:summary}.

\section{Threshold electric field}\label{sec:threshold}

The phenomenon of REs was first explored in \cite{dreicer1959,dreicer1960}; for a thermal electron population of temperature $T_e$, it was determined that an external electric field of strength $E_D = e^3 n_e \ln \Lambda /4 \pi \epsilon_0^2 T_e$ would cause the entire population to overcome friction and accelerate continuously, i.e. run away. This theory was extended in \cite{connor1975} to include relativistic effects, and the \textit{minimum} electric field needed to generate REs was found to be $E_C = E_D\times T_e/mc^2$. This threshold has been tested experimentally in many tokamaks and has been found to be a factor of at least $\sim$2-5 higher than predicted \cite{martin-solis2010,granetz2014}. The discrepancy between theory and experiment implies that other damping mechanisms on RE generation -- like synchrotron emission, bremsstrahlung, or kinetic instabilities -- can be significant compared to collisional friction. The relative importance of synchrotron radiation versus collisional friction is given by the ratio of respective timescales $\bar{\tau}_{rad} = \tau_{rad}/\tau_{coll} = 3 m n_e \ln \Lambda/2 \epsilon_0 B^2 \approx 280\, n_{20}/B^2$, where $n_{20}$ is electron density in units of $10^{20}$ m$^{-3}$ and $B$ is in Tesla. The operation of C-Mod at high magnetic fields allows exploration of $\bar{\tau}_{rad} \sim $ 2-4, values much lower than most tokamaks. As synchrotron power radiated is proportional to the square of the electron's perpendicular momentum $p_{\perp}$, the ratio of power lost per electron through synchrotron emission compared to frictional drag can simply be given by $P_{synch}/P_{coll} \approx \left( p_{\perp}/mc \right)^2/\bar{\tau}_{rad}$. As will be shown, values of $ p_{\perp}/mc \sim$~3-8 are predicted in C-Mod, meaning that synchrotron power dominates over collisional friction as a power loss mechanism.

\begin{figure}[h!]
	\centering
		\includegraphics[width=0.5\linewidth]{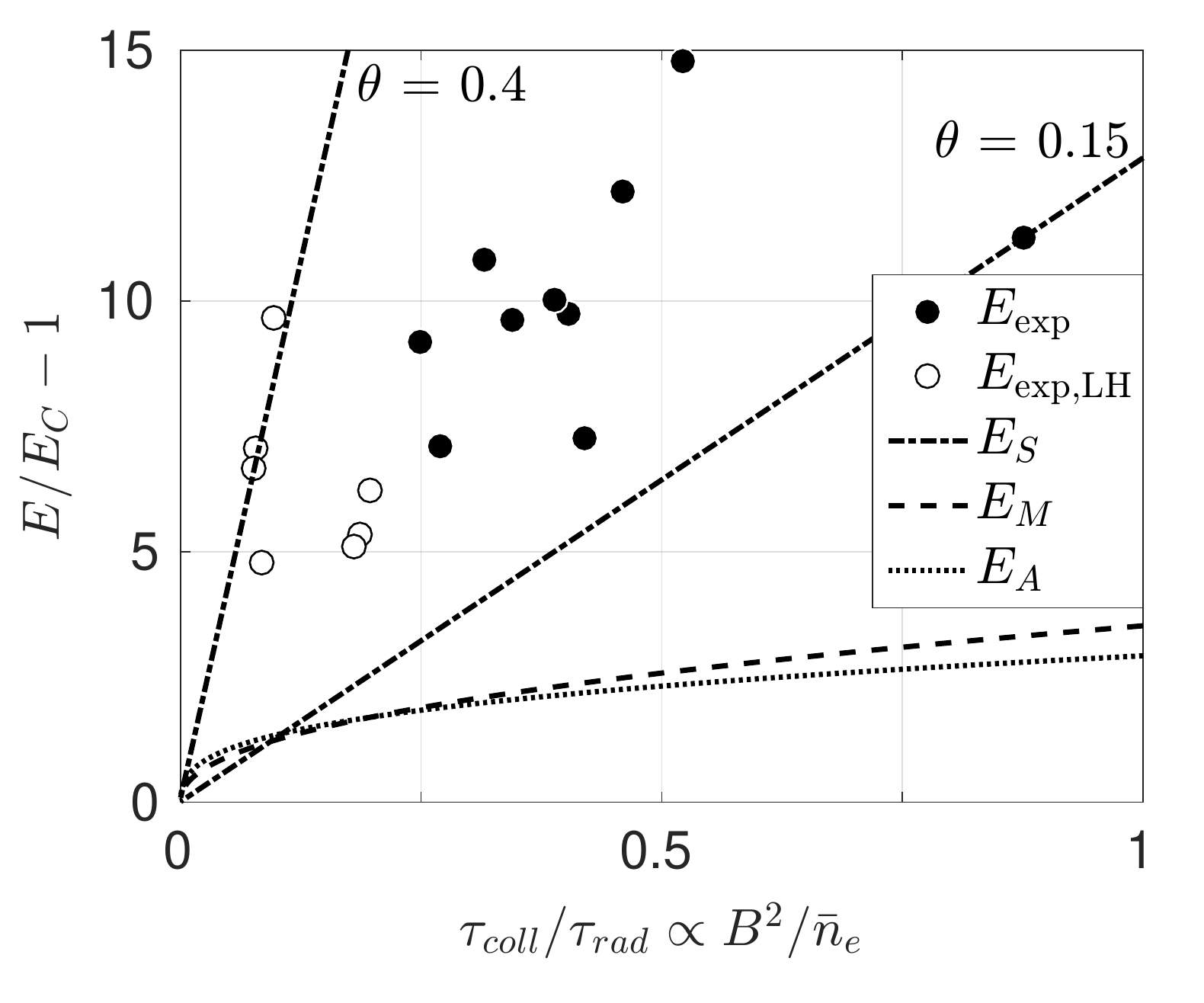}
	\caption{Ratios of the threshold to critical electric field ($E/E_C - 1$) is plotted for experimental data and from theory. The threshold fields for RE onset as measured in experiment are shown as circles, with open circles representing those discharges using LH. Theoretical predictions of the threshold field are $E_S$ (dot-dashed, (\ref{eq:E_S})), $E_M$ (dashed, (\ref{eq:E_M})), and $E_A$ (dotted, (\ref{eq:E_A})). For $E_S$, which requires inputs of energy and pitch, values of $p/mc$ = 24 ($\mathcal{E} \approx$ 12 MeV) and $\theta = p_\perp/p_\parallel$ = 0.15 and 0.4 were selected.}
	\label{fig:threshold}
\end{figure}

An analysis of RE-producing C-Mod discharges prior to 2014 is presented in \cite{granetz2014}. Since then, several RE experiments have been performed on C-Mod for a range of magnetic fields, and it is of interest to study the effect of changing $B$ on RE generation. Since not all RE discharges produce measurable synchrotron radiation, the threshold electric field is computed at the time when photo-neutron and HXR signals increase above detector noise level, i.e. at the approximate onset of RE generation. See, for example, the sharp increases of signal in figure~\ref{fig:plasmaParameters}f. The ratio of this threshold field to critical field $E_C$ is shown in figure~\ref{fig:threshold}, for REs generated during flattop $I_P$, as a function of $\tau_{coll}/\tau_{rad}$. The line-averaged electron density $\bar{n}_e$, electric field approximation $E = V/2 \pi R_0$, and Coulomb logarithm $\ln \Lambda$ = 15 were used in the calculation of $E_C$ so as to represent ``bulk" plasma paramters and be consistent with previous studies. Here $V = V_{loop} - L\,\rmd I_p/\rmd t$, where the inductance $L \approx$ 1 $\mu$H for these C-Mod conditions. Note that some discharges (open circles) used LH to encourage RE growth, especially at low $B$. These data indicate that the threshold field is $\sim$5 times higher than predicted by purely collisional theory, in agreement with previous experiments.

In recent years, several theoretical predictions of the threshold electric field have taken into account such effects as synchrotron radiation, impurities, and pitch angle scattering. In this study, three theories are compared to the experimental data shown in figure~\ref{fig:threshold}. The first comes from FTU data on the threshold field, $E_M$, required for RE \textit{suppression} \cite{martin-solis2010}; the empirical fit of the data given by equation~(3) therein is

\begin{equation}
	\frac{E_M}{E_C} \approx 1 + C(Z_{\mathrm{eff}}) \left(\bar{\tau}_{rad}\right)^{-\delta},
	\label{eq:E_M}
\end{equation}

\noindent where $C(Z_{\mathrm{eff}}) = 1.64 + 0.53 Z_{\mathrm{eff}} - 0.015 Z_{\mathrm{eff}}^2$ and $\delta = 0.45 \pm 0.03$. For this analysis, the effective charge $Z_{\mathrm{eff}}$ = 4 was chosen as a representative value, and the mean value of $\delta$ was used. In \cite{aleynikov2015}, the threshold field, $E_A$, required to \textit{sustain} REs is derived from kinetic theory; equation~(8) therein gives an analytical fit,

\begin{equation}
	\frac{E_A}{E_C} \approx 1 + \frac{(Z_{\mathrm{eff}}+1) \, \left(\bar{\tau}_{rad}\right)^{-1/2}}{\left( \frac{1}{8} + \frac{(Z_{\mathrm{eff}}+1)^2}{\bar{\tau}_{rad}} \right)^{1/6}}.
	\label{eq:E_A}
\end{equation}

\noindent However, (\ref{eq:E_A}) is only valid for $\bar{\tau}_{rad} > 5$, or $\tau_{coll}/\tau_{rad} < 0.2$ in figure~\ref{fig:threshold}. Finally, in equation~(2) of \cite{stahl2015}, the effective critical field, $E_S$, required for a particle of momentum $p = \gamma m v$ and pitch $\theta = p_\perp/p_\parallel$ to overcome collisional drag and the Abraham-Lorentz radiation reaction force is given by

\begin{equation}
	\frac{E_S}{E_C} = \sqrt{1 + \theta^2} \left( 1 + \frac{\theta^2}{1 + \theta^2}\, \frac{1 + (p/mc)^2}{\bar{\tau}_{rad}} \right),
	\label{eq:E_S}
\end{equation}

\noindent where the approximations $\beta = v/c \rightarrow 1$ and thus $\gamma = (1-\beta^2)^{-1/2} \gg 1$ were taken. Note that this definition of $\theta$ was used to be consistent with \cite{stahl2013,jaspers2001,yu2013}. The pitch \textit{angle} is therefore $\theta_p = \arctan(\theta)$.

Equations~(\ref{eq:E_M})-(\ref{eq:E_S}) are plotted as functions of 1/$\bar{\tau}_{rad}$ in figure~\ref{fig:threshold}. As is seen, experimental data are several factors greater than $E_A$ and $E_M$. For $E_S$ (\ref{eq:E_S}), a particle of energy $\mathcal{E} \approx$ 12 MeV ($p/mc$ = 24) was chosen as an estimate for the energy at which photo-neutrons (resulting from thick-target bremsstrahlung gamma rays of similar energy) would be produced. Realistic pitches of $\theta$ = 0.15 and 0.4 were selected to bound the experimental data. $E_S$ provides the best prediction of threshold electric field; however, a model which does not require a priori knowledge of the RE population is most useful and necessary for predictions of the threshold electric field of future devices. This motivates a more careful study of the threshold electric field at low $\bar{\tau}_{rad}$ (high $B$) values, as an unmitigated ITER disruption could have $\bar{\tau}_{rad} \approx$ 10. Moreover, profiles of $n_e$, $T_e$, and $E$ lead to spatial variation of $E/E_C$ -- as described in later sections -- thus obscuring the meaning of a threshold field for the bulk plasma. However, this is left to future work.

\section{Synchrotron radiation theory}\label{sec:synchrotron}

\begin{figure}[h!]
	\centering
		\includegraphics[width=0.5\linewidth]{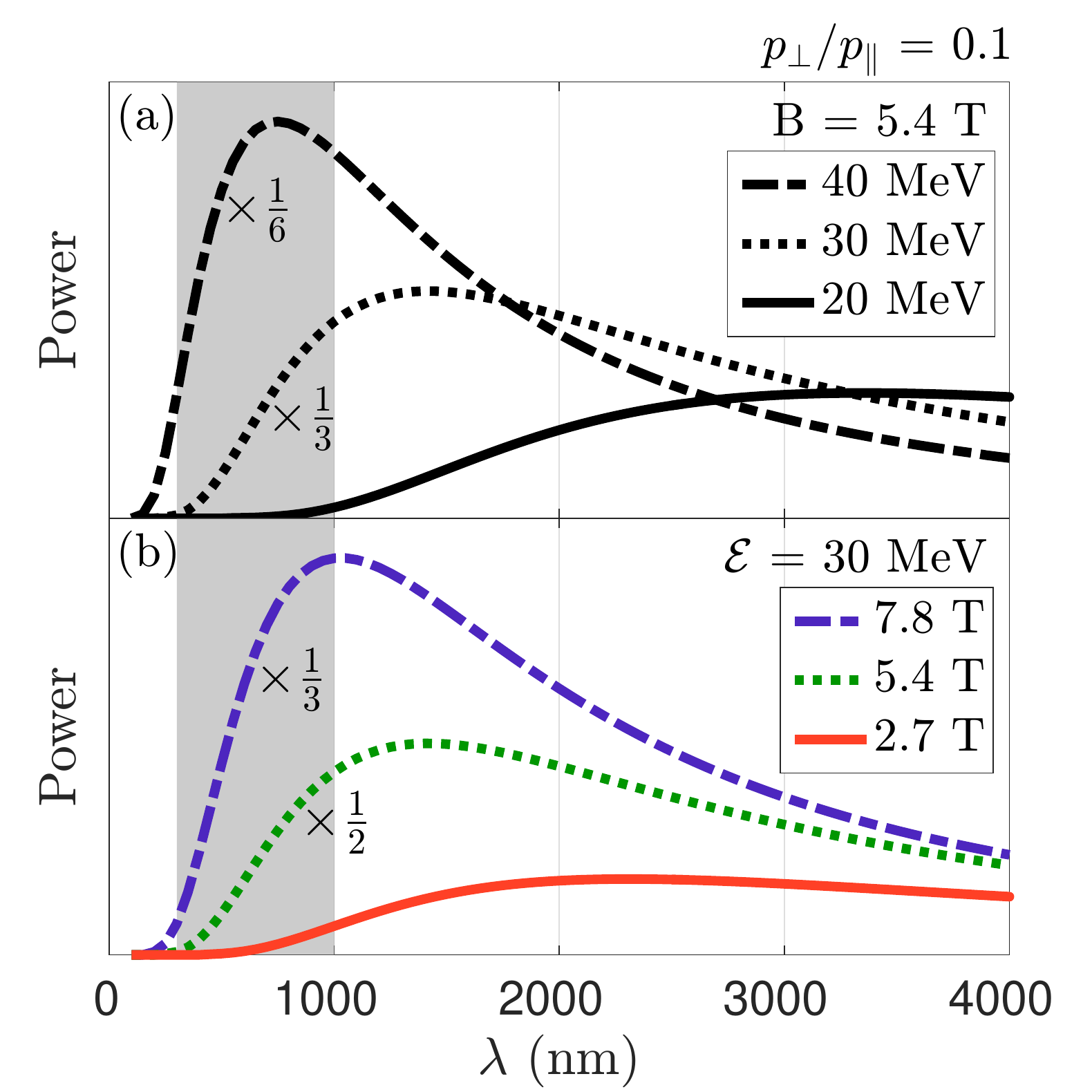}
	\caption{Synchrotron power spectra $\mathcal{P}(\lambda)$ \cite{pankratov1999,stahl2013} are plotted over the visible and near-infrared ranges for two cases: (a) At a constant $B$ = 5.4 T, the RE energy is varied among 20 MeV (solid), 30 MeV (dotted), and 40 MeV (dot-dashed); and (b) at a constant RE energy $\mathcal{E}$ = 30 MeV, $B$ is varied among 2.7 T (solid), 5.4 T (dotted), and 7.8 T (dot-dashed). The pitch $p_{\perp}/p_{\parallel} = 0.1$ is held constant for both cases, and the spectra have been re-scaled by the factors given to highlight change in spectral shape. In this study, the spectrometer wavelength range of interest ($\lambda \sim$ 300-1000 nm) is shaded.}
	\label{fig:spectraPankratov}
\end{figure}

The total power radiated by a single gyrating electron with momentum $p_\perp$ perpendicular to a magnetic field is given by the Larmor formula, $P = e^4 p_\perp^2 B^2/6\pi\epsilon_0 m^4 c^3$. In a tokamak, electrons exhibit toroidal, poloidal, and gyro-motion, all of which contribute to synchrotron emission; however in C-Mod, the smallness of the gyroradius compared to the major radius ($r_g/R_0 \approx 0.003$) means that synchrotron emission is gyro-orbit dominated. Thus, a three-fold increase in toroidal magnetic field (e.g. 2.7 to 7.8 T) will increase the power lost per electron by almost an order of magnitude. From the spectral power density of an arbitrarily-moving charge as formulated in \cite{schwinger1949}, the synchrotron power spectrum -- hereafter referred to as $\mathcal{P}(\lambda,\vec{p},t) = \rmd P_{synch}/\rmd\lambda$ (W/m) -- was calculated for a toroidal magnetic geometry in equation~(15) of \cite{pankratov1999}. The relative changes in spectra for increasing energy $\mathcal{E}$ and $B$ are shown in figure~\ref{fig:spectraPankratov}. As is seen, the total emitted power increases and peak of spectral emission shifts toward shorter wavelengths with increasing magnetic field strength and particle energy. Therefore, it might be expected that RE discharges at higher magnetic fields would increase the relative amount of blue compared to red light. However, this effect competes with more power lost by REs due to the synchrotron radiation itself, which decreases their energy and pushes the spectral peak back toward longer wavelengths. Note from figure~\ref{fig:spectraPankratov}a that RE energies of $\mathcal{E} \geq$ 40 MeV are required for the peak of emission to be in the visible wavelength range (shaded) at 5.4 T, the typical $B$-field strength of C-Mod and approximately that of ITER. 

The experimental spectral measurement of synchrotron radiation is typically of a brightness, or spectral radiance, which is a power emitted or received per unit area per unit solid angle per unit wavelength (W/m$^2$/sr/m). An early ``synthetic diagnostic" calculation for brightness was formulated in \cite{stahl2013,jaspers2001,yu2013} as

\begin{equation}
	\mathcal{B}(\lambda,\vec{p},t) \approx \frac{2 R_0}{\pi \theta_{\mathrm{eff}}(\vec{p})} n_{RE}(t) \mathcal{P}(\lambda,\vec{p},t).
	\label{eq:brightness}
\end{equation} 

\noindent Here $R_0$ is the major radius of the tokamak, and $n_{RE}$ is the density of REs. The relativistic nature of synchrotron radiation combined with toroidal and gyro-motion results in a forward-directed ``cone" of emission which must intersect the spectrometer to be detected. These geometric effects are approximated in (\ref{eq:brightness}) by the effective pitch $\theta_{\mathrm{eff}}~=~ \sqrt{\theta^2+\gamma^{-2}+(r_{ap}/r_0)^2}$, which accounts for the opening half-angle ($\theta_p~\approx~\theta~=~p_\perp/p_\parallel$) and approximate angular width ($1/\gamma$) of the emission cone (assuming $\theta^2 \ll 1$ and $\gamma\theta \gg 1$), spectrometer aperture radius $r_{ap}$, and distance $r_0$ from the aperture to tangency point of emission. This estimation assumes that the spectrometer observes only a small toroidal slice of REs since emission is approximately tangent to their toroidal trajectories. Recently, more advanced synthetic diagnostics \cite{hoppe2017,carbajal2017ppcf} have been developed and incorporate the magnetic field topology, detector geometry and spectral response function, and both momentum and spatial distributions of REs. The synthetic diagnostic SOFT has been shown to reproduce the general features of synchrotron camera images from C-Mod \cite{hoppe2017} and DIII-D \cite{hoppe2018}; it is used in this study for its synthetic spectrometer capabilities.

\section{Experimental setup}\label{sec:setup} 

To explore the impact of increased magnetic field on synchrotron power loss and RE evolution, flattop REs were generated at three magnetic fields (2.7, 5.4, and 7.8 T) in C-Mod. The experimental plasma parameters for three discharges are shown in figure~\ref{fig:plasmaParameters}. For each discharge, $n_e$ was decreased in time to achieve $E_0/E_C \geq 5$ on-axis. In addition, to encourage RE production, $I_P$ was ramped from 0.5-0.6 MA and 1.0-1.4 MA during the 2.7 and 7.8 T discharges, respectively, and LH was used for a short time from $t$ = 0.74-0.76 s during the 2.7 T discharge. These three discharges were chosen for comparison due to their similar ratios of $E_0/E_C$ on-axis. Low values of $\tau_{rad}/\tau_{coll}\sim$~2-4 for the 5.4 and 7.8 T discharges indicate the potential for synchrotron power to surpass collisional friction as the dominant power loss mechanism.

Prior to 2015, several diagnostics were available for the study of REs on C-Mod. Wide-view visible cameras allowed observation of RE spatial evolution; however, the images would often be obscured by HXR ``white-noise" or particulate build-up on camera windows. A radially-viewing energy-resolved HXR camera with 32 chords measured HXR emission from RE bremsstrahlung within the plasma up to $\sim$240 keV. Unconfined REs impacting the vessel wall also produced bremsstrahlung radiation in the form of HXRs and gammas; the latter generates photo-neutrons through ($\gamma$,n) reactions. (See  figure~\ref{fig:plasmaParameters}f-g). In addition, the Motional Stark Effect diagnostic was able to measure synchrotron radiation spectra in the visible wavelength range, but only over a narrow spectral range with width $\Delta \lambda \sim$ 5 nm.

\begin{figure}[h!]
	\centering
		\includegraphics[width=0.5\linewidth]{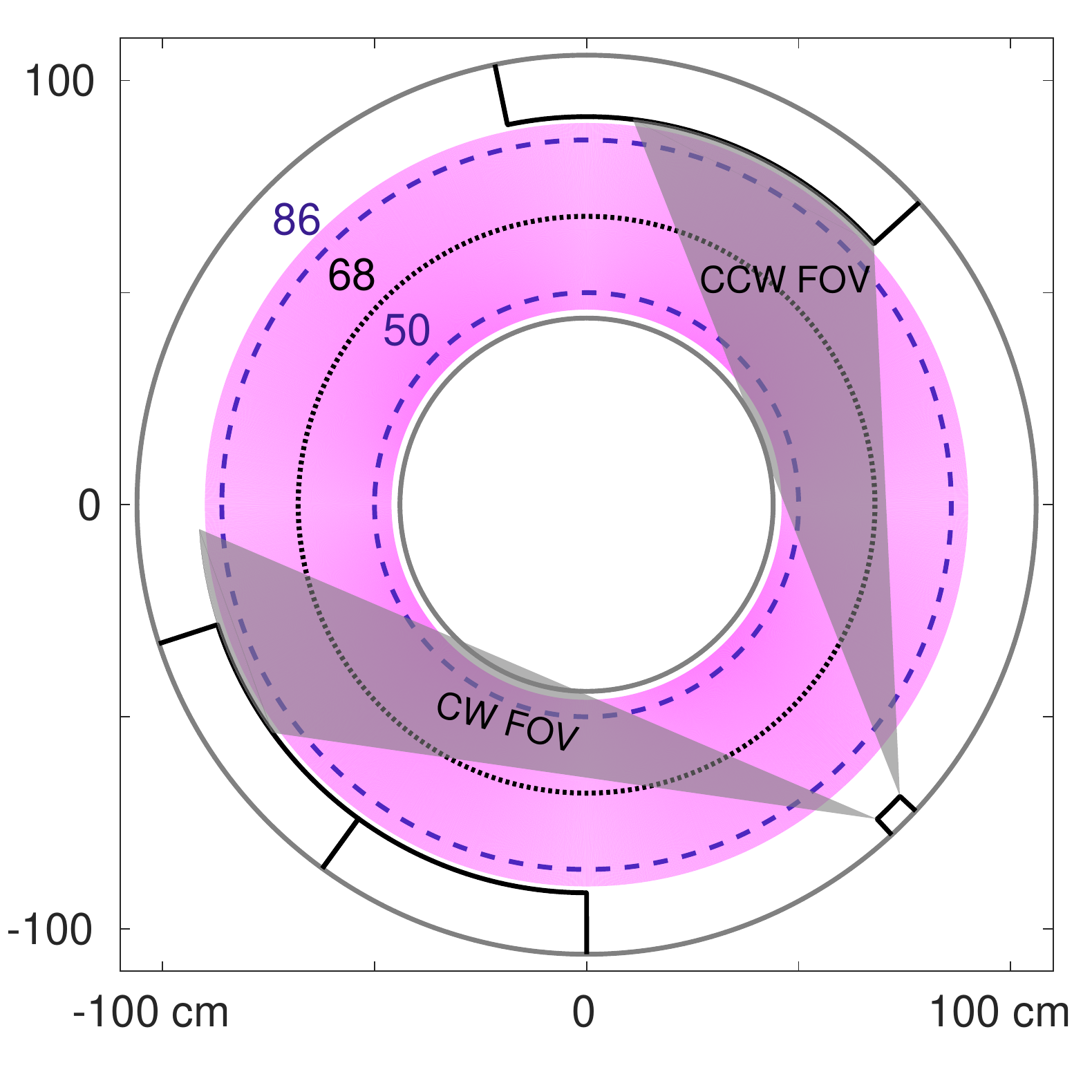}
	\caption{Clockwise (CW) and counter-clockwise (CCW) spectrometer fields-of-view (FOV) overlay a schematic midplane cross-section of the Alcator C-Mod tokamak. Major radii of 50 (dashed), 68 (magnetic axis, dotted), and 86 cm (dashed) are shown for reference.}
	\label{fig:cmod}
\end{figure}

For better spectral measurements, two visible spectrometers (see table \ref{tab:spec} for specifications) were installed on Alcator C-Mod, connected by $\sim$10 m long, 400 $\mu$m diameter silica optic fibers to apertures located approximately on the midplane viewing clockwise (CW) and counter-clockwise (CCW) directions. Opposite views allowed for the determination of the presence of REs -- as synchrotron emission is only observed from one direction--, synchrotron measurements for forward and reversed $B$ and $I_P$, and the subtraction of background light. The spectrometers were absolutely-calibrated using an in-vessel standard source in order that the measurements could be converted to absolute brightnesses. Figure~\ref{fig:cmod} shows a top-down schematic view of the tokamak midplane with the CW and CCW fields-of-view (FOV) overlaying a typical C-Mod plasma with $R_0$ = 68 cm and minor radius $a$ = 22 cm. The CW and CCW FOV total opening angles and lines-of-sight were determined from in-vessel backlighting of the optic fibers. Note that the spectrometers do not view an entire plasma poloidal cross-section, and the CW FOV does not view a tangent point on the magnetic axis. Since highly relativistic electrons will experience outward radial drifts along their helical trajectories, these combined geometric effects must be considered when analyzing synchrotron data.

\begin{table}[h!]
	\centering
	\caption{Specifications of visible spectrometers in Alcator C-Mod. Lines-of-sight are assumed horizontal (i.e. no vertical component).}
		\begin{tabular}{r c c}
		 	\hhline{===}
		 	Parameter & \multicolumn{2}{c}{Value} \\
			 \hline
		 	Wavelength range & \multicolumn{2}{c}{300-1000 nm} \\
		 	Resolution & \multicolumn{2}{c}{4 nm} \\
		 	Integration time & \multicolumn{2}{c}{1 ms} \\
		 	Data acquisition rate & \multicolumn{2}{c}{100 Hz} \\
		 	Toroidal separation & \multicolumn{2}{c}{$4.3^\circ$} \\
		 	Major radial position & \multicolumn{2}{c}{1.009 m} \\
		 	\hline
		 	Viewing direction & CW & CCW \\
		 	Full opening angle ($\alpha$) & $15.0^\circ$ & $18.8^\circ$ \\
		 	Vertical position & 3.8 cm & 2.5 cm \\
		 	Angle between radial and & \multirow{2}{*}{$-32^\circ$} & \multirow{2}{*}{$+35^\circ$} \\
		 	line-of-sight vectors & & \\
		 	\hhline{===}
		\end{tabular}
	\label{tab:spec}
\end{table}

\begin{figure*}[h!]
	\centering
	\includegraphics[width=\linewidth]{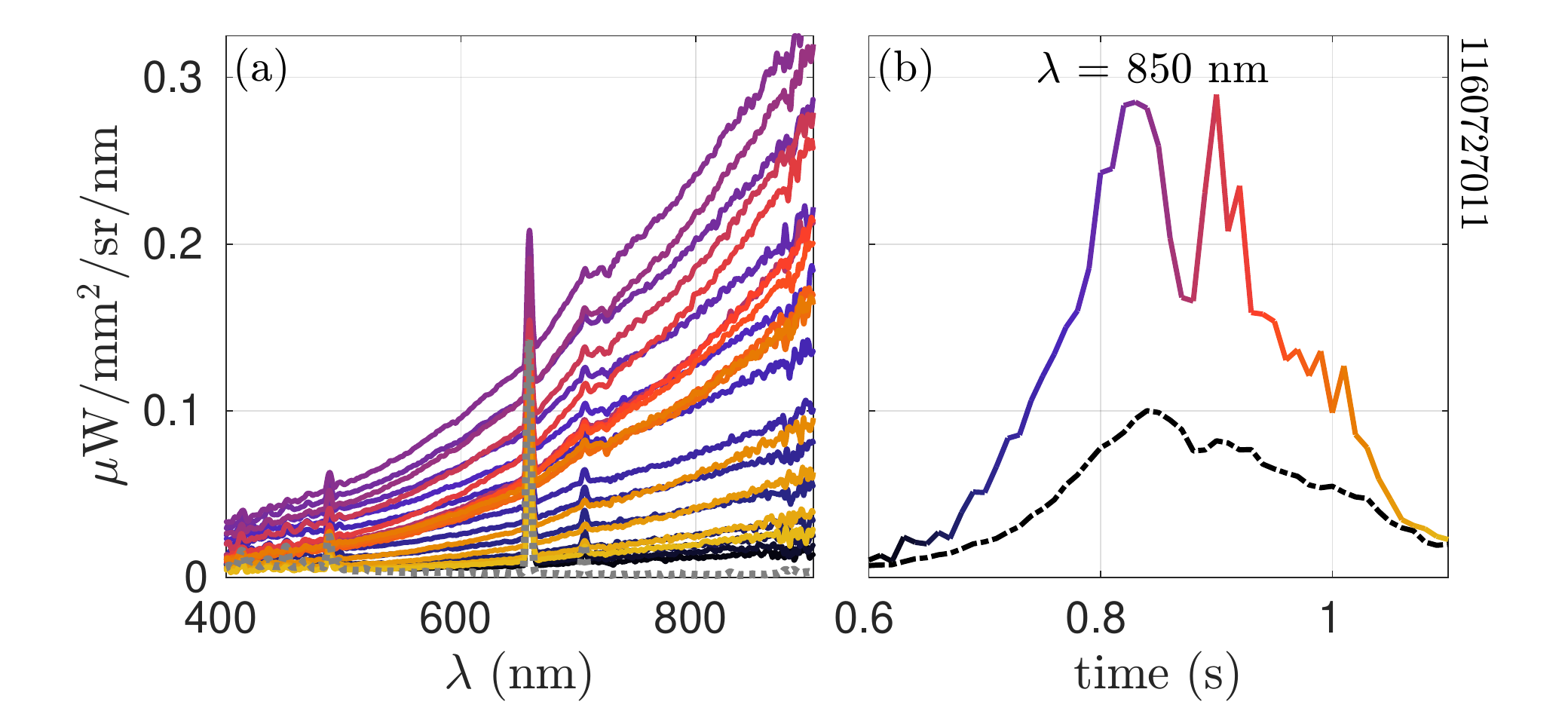}
	\caption{Synchrotron spectral data for the 7.8 T plasma discharge: (a) Absolute brightness spectra from the CW spectrometer are plotted for times between $t$ = 0.6-1.1 s at $\Delta t$ = 10 ms intervals. The dashed grey line is the time-averaged background spectrum from the opposite-direction-viewing (CCW) spectrometer. (b) Brightness at $\lambda$ = 850 nm is plotted as a function of time with the same color-coding as spectra. The HXR signal (dot-dashed, arbitrary units) from figure~\ref{fig:plasmaParameters}g is shown for comparison.}
	\label{fig:spectralData}
\end{figure*}

Absolute brightness spectra from the 7.8 T discharge are shown in figure~\ref{fig:spectralData}a, with each solid curve representing one measurement in time. The dotted line is the time-averaged background spectrum from the co-$I_P$ viewing spectrometer. The contrast between spectra clearly indicates a measurement of synchrotron radiation for the counter-$I_P$ viewing system. The absolute calibration of the spectrometers  allows qualitative comparison of the measured brightnesses with theoretical power spectra $\mathcal{P(\lambda)}$ in the visible wavelength range as shown in figure~\ref{fig:spectraPankratov}. While deuterium Balmer peaks ($\lambda \sim$ 488 and 658 nm) are visible in the absolute brightness spectra, the fiber absorption ``dips" ($\lambda \sim$ 590, 730, and 880 nm) are essentially removed by the absolute calibration. The time evolution of brightness amplitude can also yield information about RE dynamics. Figure~\ref{fig:spectralData}b shows an interesting feature of two peaks in time, at $t \sim$ 0.8 and 0.9 s, somewhat similar to the HXR signal, which occur approximately after the $I_P$ ramp shown in figure~\ref{fig:plasmaParameters}a.

\section{Approach}\label{sec:approach}

The analysis of synchrotron spectra is subtle as many factors contribute to the measured brightness, so careful interpretation of data is necessary. First attempts \cite{jaspers2001,yu2013} to determine RE parameters from spectra assumed that mono-energetic (i.e. single-energy, single-pitch) distributions of REs concentrated in the plasma core emitted synchrotron radiation directly into the spectrometer's FOV. For a highly-collimated spectrometer viewing the magnetic axis, such as that reported in \cite{yu2013}, this approximation has some merit. However, for un-collimated spectrometers, like those used in this study, the picture is much more complex: The RE population has a distribution of energies and pitches which evolves in time with changing plasma parameters. Furthermore, these momentum and density evolutions will vary on different flux surfaces as plasma parameters and magnetic field are spatially-dependent. The path of REs along helical trajectories coupled with particle drifts can also bring REs into and out of the spectrometer FOV. Solving the inverse problem, i.e. determining the spatiotemporal RE population and momentum distribution functions from time-evolving spectra, is intractable. Instead, in this paper, the forward problem is pursued by using a methodology similar to that used in \cite{esposito2003,popovic2016,esposito2017}:

\begin{enumerate}
	\item Flux-surface-averaged plasma parameters are obtained for the magnetic axis and $q$~=~1, 3/2, and 2 surfaces.
	\item A test particle model (TPM) is implemented to evolve RE momenta and densities on each surface.
	\item Drift orbits are calculated, and REs which lose confinement are removed from the population.
	\item At each time, the highest energy REs on each surface, as predicted by the TPM, are assumed to contribute most to the synchrotron spectra.
	\item The resulting evolutions of energy and density in time and space are input into SOFT to produce synthetic synchrotron spectra, which are then compared to experiment. 
\end{enumerate}  

\noindent Our approach differs from those in \cite{popovic2016,esposito2017} by including spatial and drift orbit effects, not attempting to approximate the full RE energy distribution, and utilizing SOFT. In \cite{esposito2003}, drifts and radial diffusion -- which was not explored in the present study -- were used to compare TPM energies to those measured by gamma ray spectra resulting from unconfined REs impacting PFCs.  

\subsection{Motivation}

\begin{figure*}[h!]
	\centering
	\includegraphics[width=\linewidth]{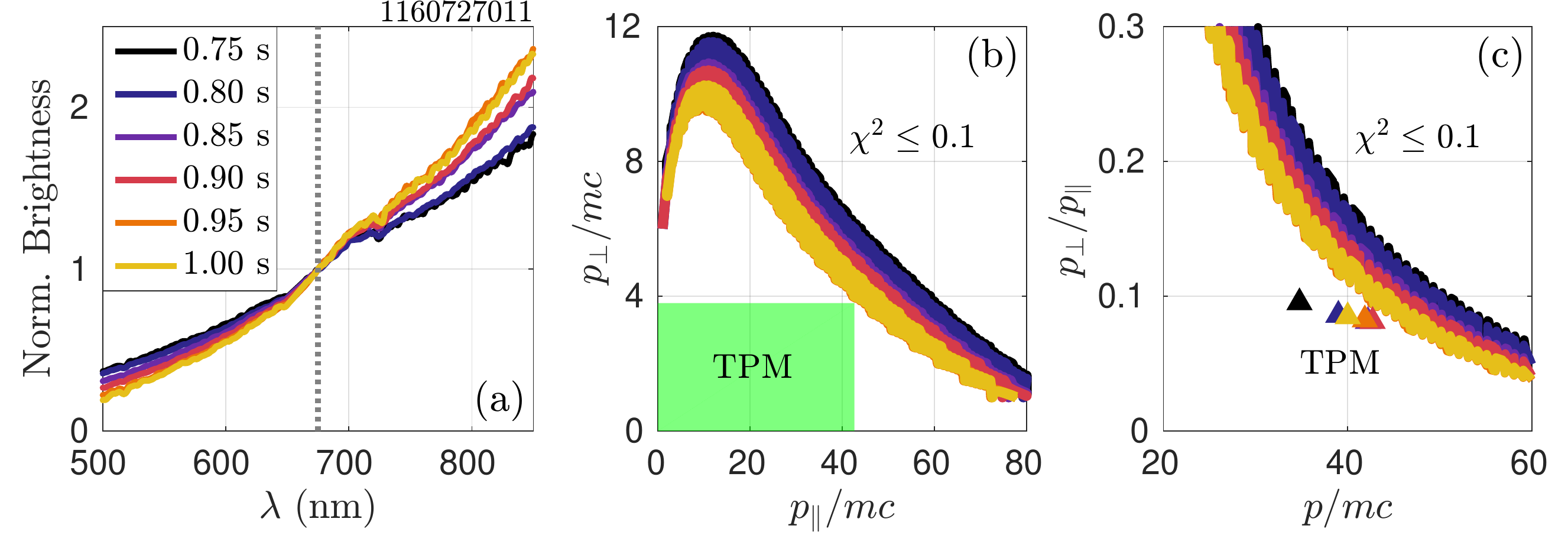}
	\caption{(a) Brightness data from figure~\ref{fig:spectralData}a is normalized at $\lambda_0$ = 675 nm (dotted). Regions of best fit ($\chi^2 \leq 0.1$) of experimental data in (a) to mono-energetic $\bar{\mathcal{B}}_{syn}$ are shown in (b) momentum space and (c) energy-pitch space. A TPM using plasma parameters from the 7.8 T discharge predicts REs confined to the shaded box in (b) and of \textit{maximum} energies and corresponding pitches as indicated by triangles in (c).}
	\label{fig:normBrightness}
\end{figure*}

When REs are generated during the flattop current of C-Mod discharges, little variation is seen in the loop voltage; thus, it is inferred that the current carried by REs is small compared to the plasma current ($I_{RE}~\ll~I_P$), making the total number of REs or RE density, $n_{RE}$, difficult to estimate. Moreover, the amplitude of synchrotron emission is a poor indicator of $n_{RE}$ as only those REs with sufficient energy ($>$10 MeV) will contribute to the visible spectra. Thus, in this analysis of synchrotron spectra, the effect of changing density is removed by normalizing the measured experimental brightness at one wavelength. The resulting normalized brightness spectra $\bar{\mathcal{B}}(\lambda,t)$ then only depend on the momenta and relative densities of REs on different flux surfaces and their contributions to the total spectra. For this study, signal-to-noise and calibration factors limit the wavelength range of interest to $\lambda$ = 500-850 nm. A normalization wavelength $\lambda_0$ = 675 nm is chosen, away from the line radiation peaks and fiber absorption features. The normalized brightness spectra are shown for several times in figure~\ref{fig:normBrightness}a. It is observed that the ``slope" of the spectra increases in time, meaning that the relative fraction of synchrotron power emitted at longer wavelengths ($\lambda > \lambda_0$) is increasing compared to shorter wavelengths ($\lambda < \lambda_0$).

In the traditional approach, these spectra would be fit using a single-particle mono-energetic approximation, as in \cite{yu2013,popovic2016,esposito2017}.\footnote{In \cite{yu2013}, the pitch was estimated from camera images of the synchrotron spot size; camera images were not available for these particular C-Mod discharges. In \cite{popovic2016,esposito2017}, a similar test particle model was used to determine the pitch, and the best-fit energy was found to be consistent with that model.} To demonstrate this, a scan in momentum space was performed from $p_\parallel/mc$ = 0-120 ($\mathcal{E}_\parallel \approx$ 0-60 MeV) and $p_\perp$ = 0-60 ($\mathcal{E}_\perp \approx$ 0-30 MeV) in increments of $\Delta p_\parallel/mc = 1$ and $\Delta p_\perp/mc = 0.5$. (Note that $p/mc \approx 2\,\mathcal{E}_{MeV}$ for electrons.) At each point, the power spectra $\mathcal{P}(\lambda,p_\parallel,p_\perp,t)$, resulting normalized synthetic brightness $\bar{\mathcal{B}}_{syn}$, and $\chi^2$ degree-of-fit of $\bar{\mathcal{B}}_{syn}$ to normalized experimental data $\bar{\mathcal{B}}_{exp}$ were calculated.


Regions of $\chi^2 \leq 0.1$ are shown in figures~\ref{fig:normBrightness}b and \ref{fig:normBrightness}c as functions of normalized parallel versus perpendicular momenta and normalized total momenta versus pitch, respectively. As is seen, there is no unique solution for this simple approach. Assuming a constant pitch $\theta = 0.1$ implies a change in energy of $\mathcal{E} \sim$ 25 to 20 MeV in time. As will be described in section~\ref{sec:TPM}, a TPM accounting for energy evolution governed by the electric field, friction, and synchrotron losses confines REs to the region of momentum space inside the shaded box ($p_\parallel/mc \leq 43$, $p_\perp/mc \leq 4$) in figure~\ref{fig:normBrightness}b. The time evolution of energy and pitch from the same TPM are represented as triangles in figure~\ref{fig:normBrightness}c. While the TPM-predicted pitch is $\theta \sim$ 0.08-0.10 -- not far from the previous assumption --, the energies are lower than the fits suggest, and the trend in time does not match. Thus, a more physically-motivated approach must be taken.

\begin{figure}[h!]
	\centering
		\includegraphics[width=0.5\linewidth]{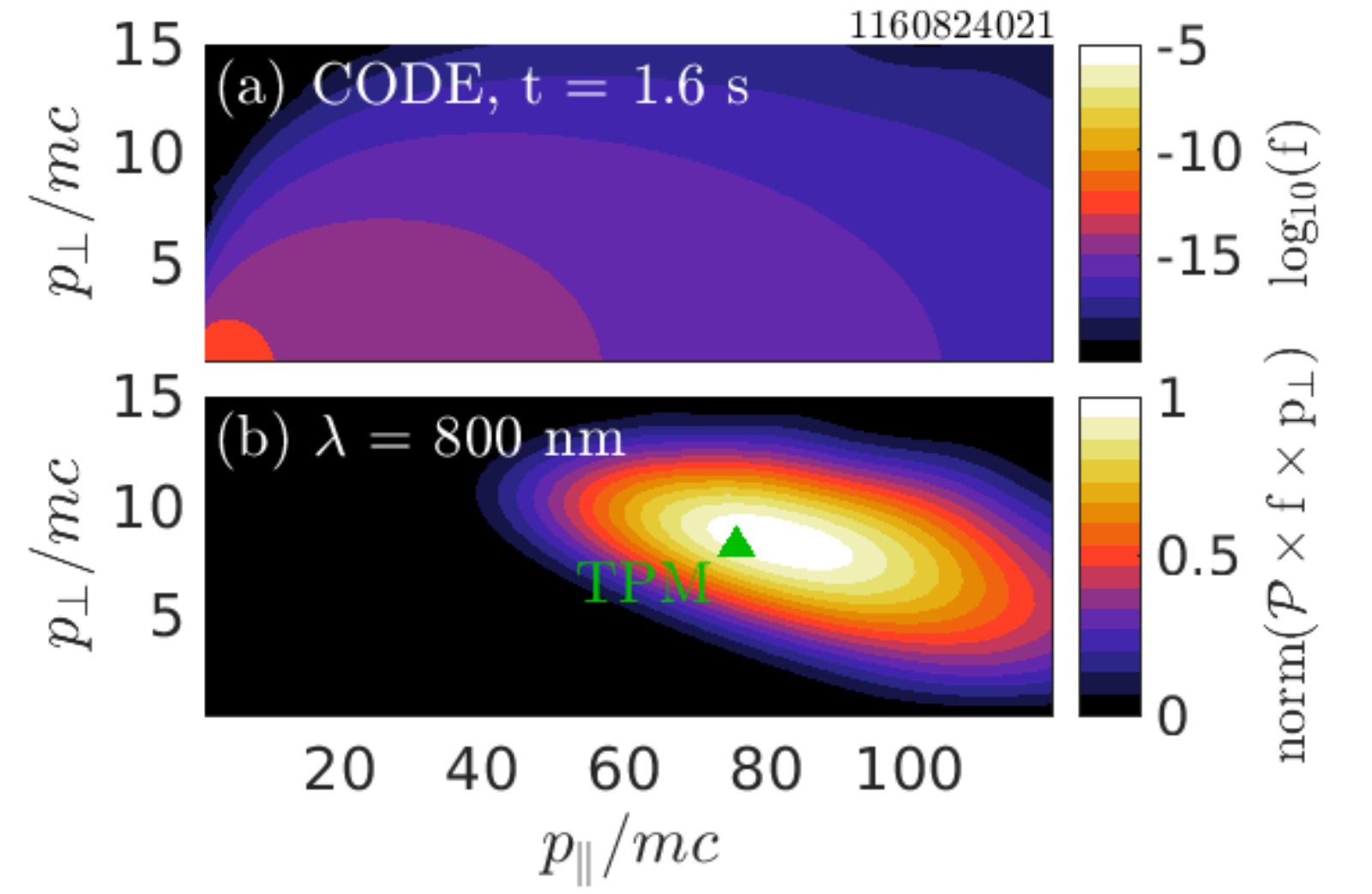}
	\caption{Contours in momentum-space of (a) the normalized momentum-space distribution function $f(p_\parallel,p_\perp)$ generated by CODE \cite{landreman2014,stahl2016} for on-axis plasma parameters of the 2.7 T discharge at $t$ = 1.6 s and (b) the normalized convolution of spectral power density $\mathcal{P}$($\lambda$ = 800 nm), $f$, and Jacobian $p_{\perp}$. The triangle in (b) represents the TPM prediction of $p_{\parallel}/mc$ and $p_{\perp}/mc$ on-axis at time $t$.}
	\label{fig:CODE}
\end{figure}

A complete analysis would include the full RE phase-space distribution function. However, an argument can be made as in \cite{hoppe2017} that because synchrotron emission is dominated by REs localized in momentum space, a single-particle approximation is sufficient to reproduce synchrotron data. Figure~\ref{fig:CODE}a shows the normalized RE energy distribution function $f(p_\parallel,p_\perp,t)$ calculated by the Fokker-Planck kinetic solver CODE \cite{landreman2014,stahl2016} using experimental plasma parameters on-axis for the 2.7 T discharge. As is seen, the distribution function is highly elongated parallel to the magnetic field as electrons are primarily accelerated in that direction. As REs are scattered to larger perpendicular momenta, they emit synchrotron radiation, lose energy, and repeat the cycle. A convolution of $f$, $\mathcal{P}$, and Jacobian $p_{\perp}$ determines the region of momentum space in which most synchrotron power is radiated.\footnote{In the guiding-center picture, gyromotion is reduced from three to two dimensions in momentum space ($p_{\parallel}$,$p_{\perp}$); thus, the Jacobian $p_{\perp}$ is required to compute the cylindrical volumetric element, $p_{\perp}\rmd p_{\parallel} \rmd p_{\perp} \rmd \theta$, for integration of the total synchrotron power.} This normalized convolution, evaluated at $\lambda$ = 800 nm, is shown in figure~\ref{fig:CODE}b. At the time shown, the peak emission results from REs with average momenta $p_\parallel/mc \sim$ 81 and $p_\perp/mc \sim$ 8. (Note that the peak location is relatively insensitive to $\lambda$ in the visible range.) The TPM, to be described, predicts a single-particle on-axis momentum of $p_\parallel/mc \sim$ 76 and $p_\perp/mc \sim$ 8, indicated by the triangle. Since parameter scans and iteration of the TPM were more feasible, i.e. less computationally-intensive, than the full kinetic solvers, a TPM was pursued. Moreover, as will be described in section \ref{sec:27T}, spectra produced using the TPM match experimental data as well or even better than those generated using CODE.

\subsection{Test particle model}
\label{sec:TPM}

Together, the intractability of solving the inverse problem and localization of synchrotron emission in momentum space motivate the use of a test particle model (TPM) to solve the forward problem: Given time-dependent background plasma parameters, let an electron of sufficiently high energy evolve in time and momentum space according to the set of coupled momentum equations \cite{martin-solis1998}

\begin{equation}
	\label{eq:ms1}
	\frac{\rmd p_{\parallel}}{\rmd t} = eE_{\parallel} - \frac{n_e e^4 m_e \ln \Lambda}{4 \pi \epsilon_0^2} \gamma \left(Z_{\mathrm{eff}} + 1 + \gamma\right) \frac{p_{\parallel}}{p^3} - \frac{e^2}{6 \pi \epsilon_0} \beta^3 \gamma^4 \langle R^{-2} \rangle \frac{p_{\parallel}}{p}
\end{equation}


\noindent and

\begin{equation}
	\frac{\rmd p}{\rmd t} = eE_{\parallel} \frac{p_{\parallel}}{p} - \frac{n_e e^4 m_e \ln \Lambda}{4 \pi \epsilon_0^2} \frac{\gamma^2}{p^2} - \frac{e^2}{6 \pi \epsilon_0} \beta^3 \gamma^4 \langle R^{-2} \rangle.
	\label{eq:ms2}
\end{equation}


\noindent Here, $E$, $n_e$, $Z_{\mathrm{eff}}$, and $\ln\Lambda$ are the time-dependent background electric field, electron density, effective charge, and Coulomb logarithm of the plasma, respectively; $\beta = v/c$ is the normalized velocity; $\gamma = (1-\beta^2)^{-1/2}$ is the relativistic Lorentz factor; and $\langle R^{-2} \rangle$ denotes the gyro-averaged radius of curvature, including both toroidal and cyclotron motions. The first, second, and third terms of (\ref{eq:ms1})-(\ref{eq:ms2}) represent the toroidal electric force, collisional drag, and synchrotron radiation reaction force, respectively. (See equation (7) of \cite{martin-solis1998} for the differential equation for $\rmd p_{\perp}/\rmd t$, which can be derived from (\ref{eq:ms1}) and (\ref{eq:ms2}).)

In addition, a model for RE density evolution can be used to predict the number of REs generated on each surface, thereby constructing an ad hoc profile and estimating relative contributions to the total synchrotron emission. This work assumes an evolution of the form

\begin{equation}
	\frac{\rmd n_{RE}}{\rmd t} = \Gamma_{\mathrm{lin}} n_e + \Gamma_{\mathrm{exp}} n_{RE}.
	\label{eq:nRE}
\end{equation}

\noindent The first term represents primary, or Dreicer, generation which dominates early in time as electrons accelerate from the tail of the thermal population. The full relativistic formulation of the linear growth term $\Gamma_{\mathrm{lin}} n_e$ is given by equation~(63) in \cite{connor1975}, and the constant of order unity therein is chosen to be 1. In general, it is found that the Dreicer growth rate increases with the temperature in the ratio $E_D/E~\propto~T_e/E$ and decreases with $Z_{\mathrm{eff}}$. 

The second term in (\ref{eq:nRE}) represents the secondary avalanching mechanism, i.e. exponential growth later in RE evolution during which thermal electrons are ``knocked" into the runaway regime through collisions with existing REs. The formulation of $\Gamma_{\mathrm{exp}}$ implemented in this work is given by equation~(18) in \cite{rosenbluth1997} with the parameter $\gamma$ therein set to 1 in the large aspect ratio limit. The growth rate $\Gamma_{\mathrm{exp}}$ is found to increase with $E/E_C$ and also decrease with $Z_{\mathrm{eff}}$.

Because REs can be generated at any time throughout the discharge, new particles are initiated using the TPM starting at $t$ = 60 ms (corresponding to the first EFIT \cite{lao1985} time) and in $\Delta t$ = 10 ms intervals. This early start time is motivated by spikes in HXR signals early in the three discharges (figure~\ref{fig:plasmaParameters}f). Regardless of the time of initiation, most REs generated during the first $\sim$0.5-1.0 s reach an approximately steady-state momentum by the time of observed synchrotron emission. In general, REs initiated earliest in time gain the highest energies and densities; however, they can lose confinement, as described in the next section (\ref{sec:drifts}), and thus lower energy runaways generated at later times can become important (see figure \ref{fig:tpm}). For simplicity, the initial momenta were chosen to be $p_{\parallel,0}/mc = 1$ (as done in \cite{esposito2003}) and $p_{\perp,0}/mc = 0.001$ which are representative of an electron above the threshold velocity $v_c$ along the magnetic field, but not highly relativistic. In addition, the initial density was $n_{RE,0} = 1$ m$^{-3}$. For both momentum and density evolutions, the TPM is insensitive to initial conditions.

\subsection{Spatial profiles and drifts}\label{sec:drifts}

Since our spectrometers make volume-integrated measurements over significant cross-sections of the plasma (figure~\ref{fig:cmod}), it is important to account for the energy and density evolutions of REs on different flux surfaces, as well as the spatial variation of the magnetic field. The spatial profile of runaways is approximated by calculating the evolutions for four locations: the magnetic axis and on the $q$~=~1, 3/2, and 2 surfaces. Surface-averaged plasma parameters were input into the TPM: The $q$-profile and electric field $E = (\partial \psi_p/\partial t)/(2\pi R)$ were determined from EFIT magnetic reconstructions \cite{lao1985}, where $\psi_p$ is the poloidal magnetic flux. Thomson scattering profiles of $n_e$ and $T_e$ on the outer midplane were used and assumed to be flux surface functions. Because $B$ and $E$ both decrease as $\sim$1/$R$, particles on helical trajectories necessarily experience varying electric and magnetic fields; it was assumed that all REs are passing particles with constant speeds to calculate the surface-averages. Due to stray synchrotron light in the diagnostic measuring effective charge, values of $Z_{\mathrm{eff}}$ could not be accurately measured for these discharges. Therefore, a scan in $Z_{\mathrm{eff}}$ was performed from 1-4 and was held constant in time and space in each TPM simulation. (Note that $Z_{\mathrm{eff}} \sim$ 3-4 is common for low density C-Mod discharges.) In this paper, only data are shown for those values of $Z_{\mathrm{eff}}$ which best reproduced experimental results. 

Highly relativistic particles also experience large particle drifts. Thus, REs of sufficient energy can lose confinement, especially if they are generated on outer flux surfaces. From \cite{knoepfel1979,myra1992,esposito1996,martin-solis1999,esposito2003}, the drift orbit radius can be expressed as $r_d \approx \frac{p_\parallel}{e\, B_p(r)} \frac{r}{R_0}$, where $B_p(r)$ is the poloidal magnetic field, $r$ is the minor radial coordinate, and $\theta^2 \! \ll \! 1$ is assumed. For a current density profile of the form $J(r) \propto 1 - (r/a)^n$, $r_d$ can be written as

\begin{equation}
	r_d = \frac{\pi}{\mu_0 e} \frac{p_\parallel}{I_P} \frac{a^2}{R_0} \left\{ \frac{1}{2} + \frac{1}{n} \left[ 1 - \left(\frac{r}{a} \right)^n \right] \right\}^{-1}.
	\label{eq:drift}
\end{equation}

\noindent Equation (\ref{eq:drift}) can be rearranged to give the maximum parallel momentum above which a particle at radius $r$ will leave the plasma, i.e. $r + r_d > a$. (The gyroradius is not included here, but is typically sub-centimeter for small pitch angles.) For a parabolic ($n = 2$) current profile and $R_0/a \approx 3.1$ for C-Mod, REs will lose confinement when

\begin{equation}
	\frac{p_\parallel}{mc} \geq 724\; I_{MA} \left(1-\frac{r}{a}\right)\left[1-\frac{1}{2}\left(\frac{r}{a}\right)^2\right],
	\label{eq:pmax}
\end{equation} 

\noindent where $I_{MA}$ is $I_P$ expressed in MA. This implies that the maximum energy of confined REs generated on-axis in a typical 1 MA discharge is $\sim$360 MeV!  Note that this does not include radial transport mechanisms, such as magnetic field perturbations.

\begin{figure}[h!]
	\centering
		\includegraphics[width=0.5\linewidth]{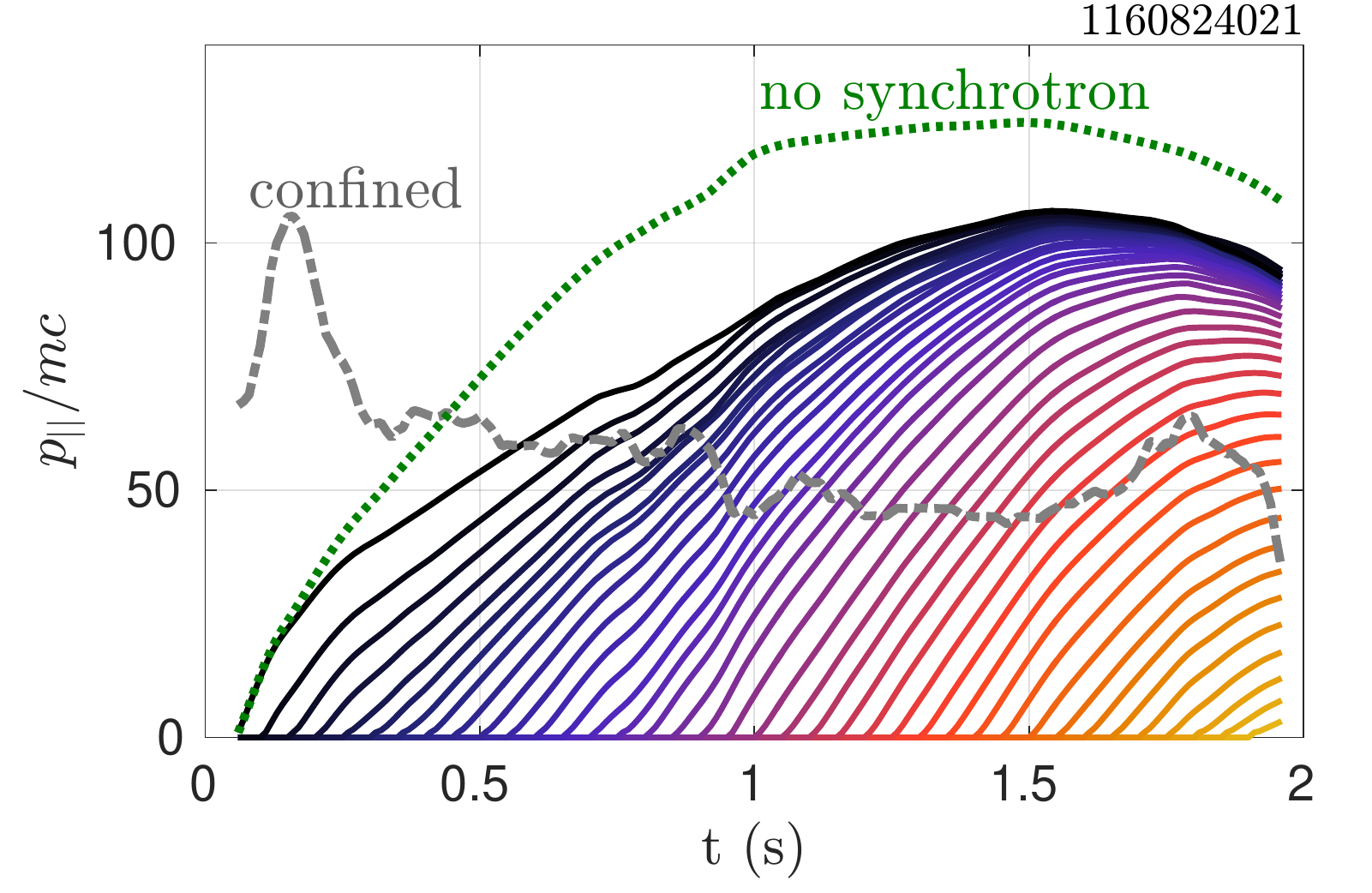}
	\caption{Time evolution of parallel momenta (solid lines) for particles initiated every 50 ms on the $q$~=~2 surface of the 2.7 T discharge. The maximum parallel momenta for RE confinement and that with negligible synchrotron radiation ($B = 10^{-4}$~T = 1~G) are the labeled dot-dashed and dotted curves, respectively. ($Z_{\mathrm{eff}} = 4$)}
	\label{fig:tpm}
\end{figure}

An illustration of the TPM and drift orbit calculations is shown in figure~\ref{fig:tpm}. Here the TPM was simulated for plasma parameters on the $q$~=~2 surface of the 2.7 T discharge. Each solid curve represents the momentum evolution (from (\ref{eq:ms1})-(\ref{eq:ms2})) of a particle generated at a different time within the discharge. If drift orbit losses are not considered, it can be seen that REs produced in the first $\sim$1 s of the discharge reach a momentum-space attractor at $p_{\parallel}/mc \approx 100$ ($\sim$50 MeV) around $t$ = 1.5 s as the electric, friction, and synchrotron radiation reaction forces balance. The importance of synchrotron power loss is highlighted by the dotted curve for which the magnetic field was reduced to $B$ = 10$^{-4}$ T, and REs reached a maximum energy of $\sim$60 MeV, 10 MeV higher than particles subject to synchrotron power loss. The maximum allowable momentum beyond which REs would drift out of the plasma is calculated from (\ref{eq:pmax}) and is shown by the dot-dashed line. This indicates that REs would lose confinement early in the shot, around $t$ = 0.6 s, and therefore the maximum energy expected is actually $\sim$25-30 MeV ($p_{\parallel}/mc \sim$ 50-60). 

\subsection{Implementation of SOFT}

The synthetic spectrometer capabilities of the \textit{Synchrotron-detecting Orbit Following Toolkit}, SOFT, \cite{hoppe2017} were utilized in this work. Inputs to SOFT include the magnetic geometry from EFIT \cite{lao1985}; spectrometer specifications (e.g. position, orientation, viewing angle, and spectral response) provided in table \ref{tab:spec}; total particle momenta and pitch angles from the TPM or momentum space distribution functions from CODE; and midplane major radial locations of REs. The synthetic diagnostic works as follows: REs are initiated on the outer midplane with prescribed momenta and radial positions. These particles execute guiding center motion along magnetic field lines, and their pitch angles evolve by conservation of magnetic moment. Drift orbit effects are not yet fully implemented in the simulation and thus are not utilized here. The full angular distribution was used for the synchrotron power computation. (Although the simplified cone model -- with opening half-angle $\theta_p$ and negligible angular width -- was found to work just as well.) If emitted radiation hits the spectrometer aperture, its energy is recorded, and a spectrum is produced.

In this work, separate SOFT simulations were performed for each flux surface, $q_i$, and time of interest. In each simulation, the same number of particles were distributed uniformly in a radial band of width $\Delta R$ = 1 cm centered at the outer midplane major radius corresponding to the flux surface. The resulting spectra were scaled by the total number of REs expected on each surface, calculated using the RE density and cylindrical plasma approximation:

\begin{equation}
	N_{RE}(q_i,t) \approx n_{RE}(q_i,t) \times 2\pi R_0 \times 2 \pi \left( R(q_i,t) - R_0 \right) \Delta R.
\end{equation}


\noindent The total brightness is then simply 

\begin{equation}
	\mathcal{B}_{SOFT,tot}(t) = \sum\limits_i N_{RE}(q_i,t)\, \mathcal{B}_{SOFT}(q_i,t),
\end{equation}

\noindent which can be normalized and compared to experiment.

\section{Experimental analysis}\label{sec:analysis}

The following subsections detail the application of the prescribed methodology to the three plasma discharges of interest with magnetic fields of $B_0$ = 2.7, 5.4, and 7.8 T. The organization is as follows: Plasma parameters $n_e$, $T_e$, $E/E_C$, and outer midplane major radial position $R$ are shown for the magnetic axis and $q$~=~1, 3/2, and 2 surfaces in figures~\ref{fig:plasmaParams27T}, \ref{fig:plasmaParams54T}, and \ref{fig:plasmaParams78T} as subplots (a)-(d), respectively. In the same figures, TPM results of $p/mc$, $p_\perp/p_\parallel$, and $n_{RE}$ are shown in subplots (f)-(h) for each surface. Note that drift orbits are accounted for in the TPM calculation and are also shown as dotted lines in subplot (d). The time evolution of measured synchrotron brightness is shown in subplot (e) for qualitative comparisons with TPM parameters; vertical lines indicate the times at which synthetic spectra are produced.

In figures~\ref{fig:normBrightness27T}, \ref{fig:normBrightness54T}, and \ref{fig:normBrightness78T}, the synthetic data are compared to experimental spectra for the 2.7, 5.4, and 7.8 T discharges, respectively. In each figure, subplot (a) shows the experimentally-measured normalized brightness spectra. The synthetic spectra shown in the accompanying subplots (b) and (c) are calculated using (b) TPM data only on-axis input into $\mathcal{B}$ (\ref{eq:brightness}) and (c) TPM data from all surfaces input into SOFT. Subplot (d) is different for each discharge. Finally, the $\chi^2$ degree-of-fit is shown in figure~\ref{fig:X2} and discussed in section~\ref{sec:discussion}.

\subsection{$B_0$ = 2.7 T}\label{sec:27T}

\begin{figure}[h!]
	\centering
		\includegraphics[width=0.75\linewidth]{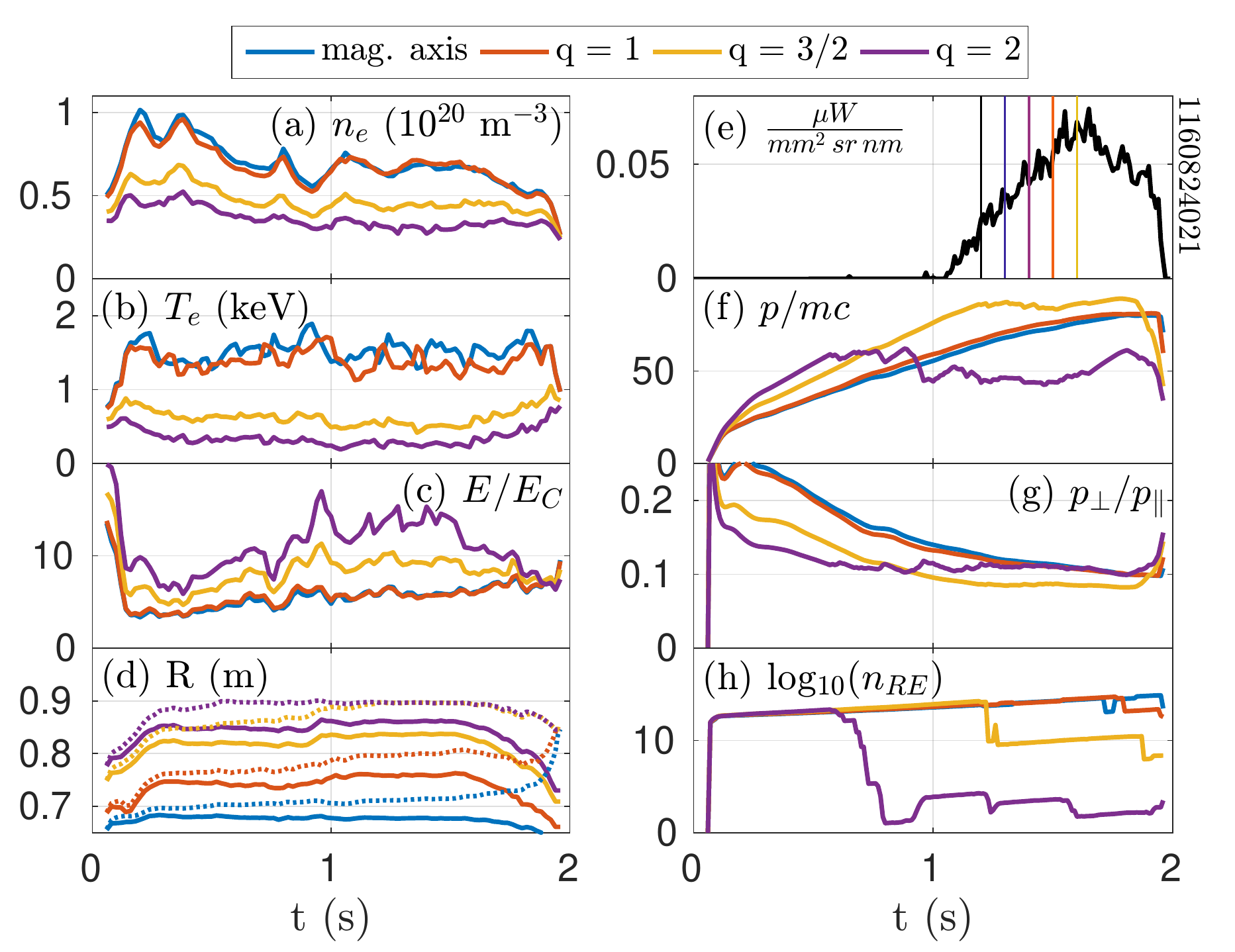}
	\caption{For the 2.7 T discharge, parameters on the magnetic axis and $q$~=~1, 3/2, and 2 surfaces: (a) electron densities, (b) electron temperatures, (c) ratios of electric to critical fields, and (d) outer midplane major radii with (dotted) and without (solid) including drifts; (e) measured synchrotron brightness (at $\lambda$ = 850 nm) with times of interest corresponding to figure~\ref{fig:normBrightness27T} indicated as vertical lines; and TPM results for RE (f) total normalized momenta, (g) pitches, and (h) densities. ($Z_{\mathrm{eff}} = 4$)}
	\label{fig:plasmaParams27T}
\end{figure}

During the 2.7 T discharge, $n_e$ was decreased in time (figure~\ref{fig:plasmaParams27T}a) and $I_p$ was ramped from 0.5-0.6 MA over $t$ = 0.9-1.0 s (figure~\ref{fig:plasmaParameters}a) to encourage RE generation. (LH current drive was also used to produce a seed population, but was not modeled explicitly.) The resulting increase in $E/E_C$ due to the $I_P$ ramp can be seen during this time interval in figure~\ref{fig:plasmaParams27T}c, especially for the outer surfaces, $q$~=~3/2 and 2. Higher values of $E/E_C$ on the outer flux surfaces lead to higher predicted RE momenta (figure~\ref{fig:plasmaParams27T}f) early in the plasma discharge. However, the resulting drift orbits (figure~\ref{fig:plasmaParams27T}d) approach the plasma boundary ($R \approx$ 0.9 m), and REs lose confinement at $t \sim$ 0.6 and 1.2 s for the $q$~=~2 and 3/2 surfaces, respectively; these are seen as decreases in energy (figure~\ref{fig:plasmaParams27T}f) and density (figure~\ref{fig:plasmaParams27T}h). The former time, $t \approx$ 0.6 s, is approximately the time at which photo-neutron and HXR signals rise in figure~\ref{fig:plasmaParameters}f-g. While it is predicted that the highest energy REs are on the $q$~=~3/2 surface, $n_{RE}$ there is far lower than on-axis. Thus, it is expected that REs at the core will dominate the synchrotron spectra.

\begin{figure}[h!]
	\centering
		\includegraphics[width=\linewidth]{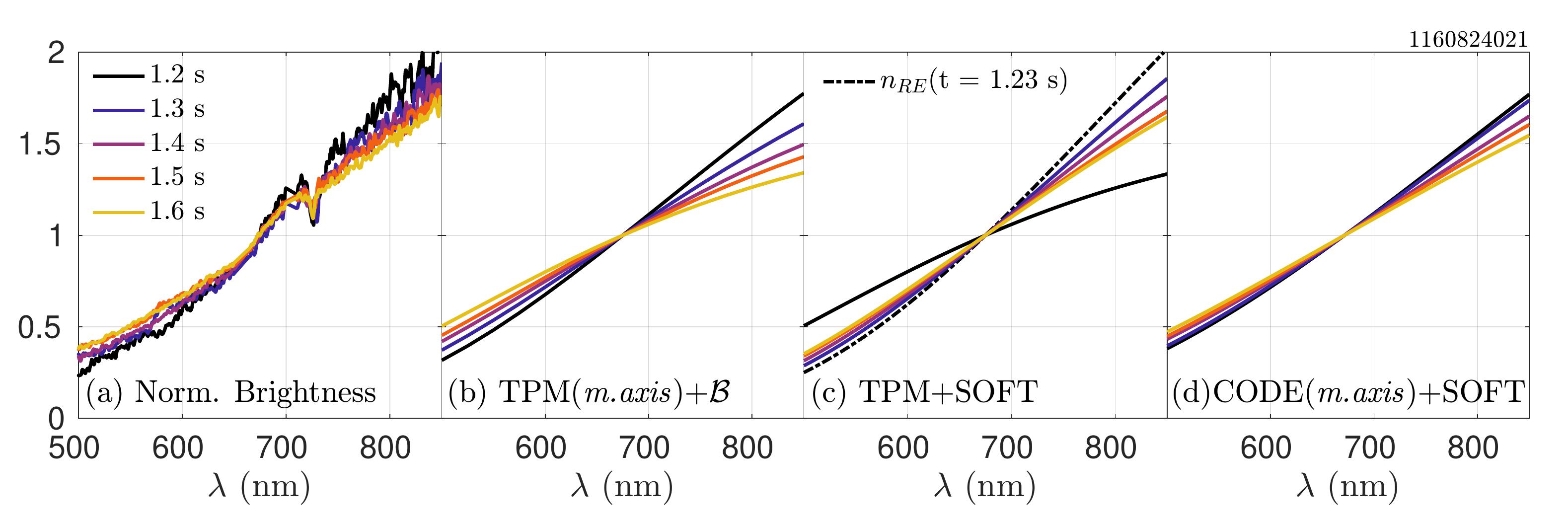}
	\caption{For five times during the 2.7 T discharge: comparisons of (a) experimental normalized brightness spectra to synthetic data from (b) on-axis TPM data input into $\mathcal{B}$ (\ref{eq:brightness}), (c) TPM data from all surfaces input into SOFT, and (d) on-axis CODE distribution functions input into SOFT. The dot-dashed line in (c) uses RE densities at $t$ = 1.23 s. Note that the full angular distribution was used for SOFT in (b), while the cone model approximation was used in (d).}
	\label{fig:normBrightness27T}
\end{figure}

Synchrotron emission was measured by the CCW spectrometer (see figure~\ref{fig:cmod}) for the 2.7 T discharge. The experimental normalized brightnesses at five times, as indicated in figure~\ref{fig:plasmaParams27T}e, are compared to synthetic spectra in figure~\ref{fig:normBrightness27T}. Considering REs only on-axis, and using the brightness formula $\mathcal{B}$ (\ref{eq:brightness}), produces spectra similar to experiment, as seen in figure~\ref{fig:normBrightness27T}b. This is consistent with measured synchrotron emission dominated by REs in the core. Combining TPM data from all surfaces and inputting it into SOFT produces synthetic spectra most similar to experiment, except at $t$ = 1.2 s (figure~\ref{fig:normBrightness27T}c). This discrepancy is due to a significant contribution of synchrotron emission from high energy REs on the $q$~=~3/2 surface before they lose confinement. However, at $t$ = 1.23 s, slightly after the loss of these REs on the $q$~=~3/2 surface, the model spectra match experiment. This could indicate that REs on the $q$~=~3/2 surface are lost $\sim$30 ms earlier than the TPM and drift orbit calculations predict, perhaps through an additional transport mechanism. Such an explanation seems physically plausible. Finally, the full distribution function from CODE, using plasma parameters only on-axis, was input into SOFT to produce synthetic spectra (figure \ref{fig:normBrightness27T}d). As is seen, the TPM+SOFT synthetic spectra match the experimental spectra as well as the CODE+SOFT data, further motivating the use of the TPM approximation.

In addition, note that drift orbits of $\sim$4 cm (figure~\ref{fig:plasmaParams27T}d) are predicted for REs on the magnetic axis and $q$~=~1 surfaces at the time of observed synchrotron emission. This implies that the trajectories of REs ``on-axis" are actually at the boundary of the CCW spectrometer FOV (figure~\ref{fig:cmod}). The SOFT synthetic spectra in figure~\ref{fig:normBrightness27T}c-d do not include drift effects. However, since both inner surfaces would still be seen by the CCW spectrometer even with drifts, the resulting synthetic spectra would likely be unchanged.

\subsection{$B_0$ = 5.4 T}

\begin{figure}[h!]
	\centering
		\includegraphics[width=0.75\linewidth]{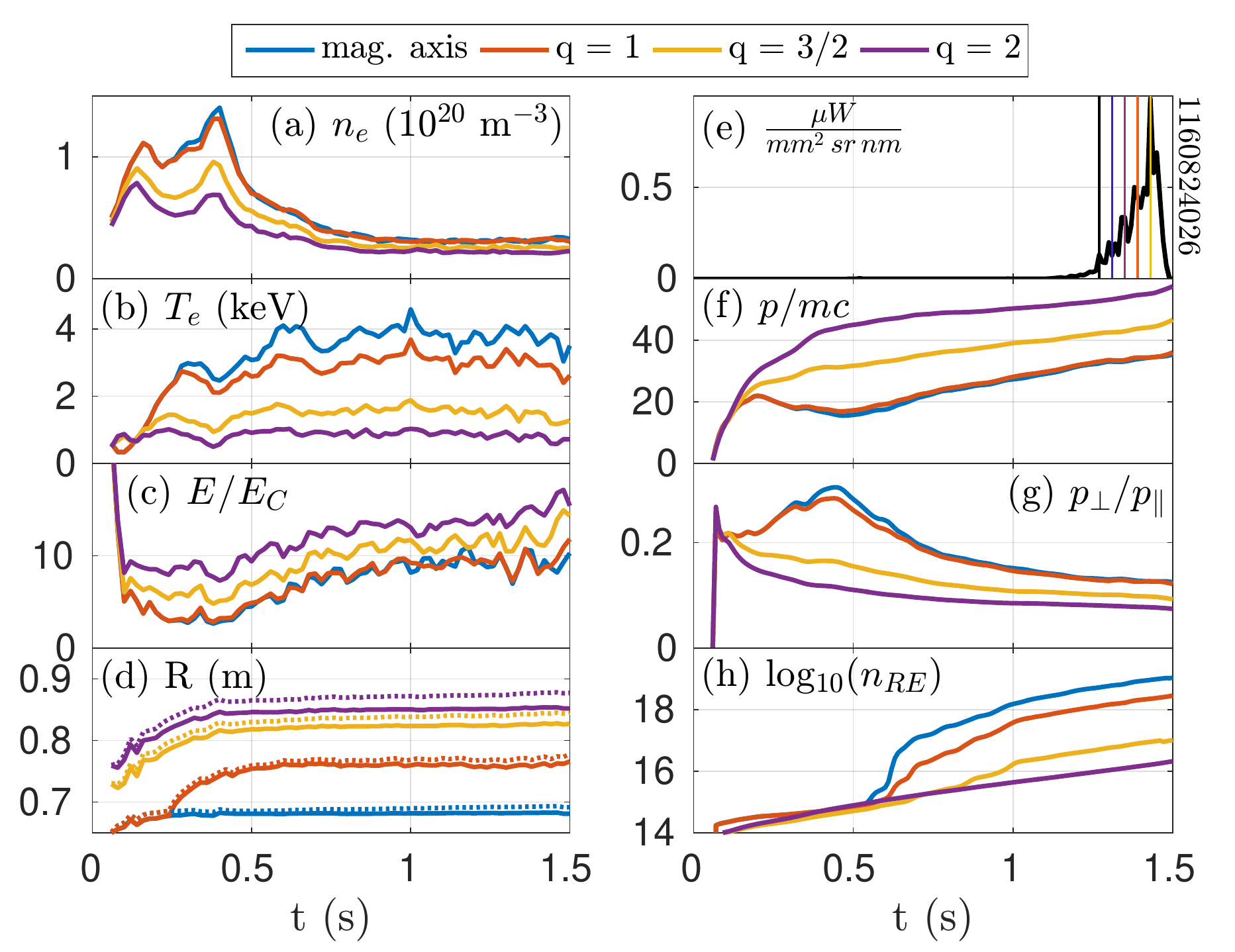}
	\caption{For the 5.4 T discharge, parameters on the magnetic axis and $q$~=~1, 3/2, and 2 surfaces: (a) electron densities, (b) electron temperatures, (c) ratios of electric to critical fields, and (d) outer midplane major radii with (dotted) and without (solid) including drifts; (e) measured synchrotron brightness (at $\lambda$ = 850 nm) with times of interest corresponding to figure~\ref{fig:normBrightness54T} indicated as vertical lines; and TPM results for RE (f) total normalized momenta, (g) pitches, and (h) densities. ($Z_{\mathrm{eff}} = 3.5$)}
	\label{fig:plasmaParams54T}
\end{figure}

Similar to the 2.7 T discharge, $n_e$ was decreased in time during the 5.4 T discharge to encourage the growth of REs (see figure~\ref{fig:plasmaParams54T}a). The surface-averaged values of $E/E_C$ are consistently higher on the $q$~=~3/2 and 2 surfaces than near the center of the plasma, which lead to higher energies predicted on outer surfaces, by as much as $\sim$5-10 MeV. However, these RE energies are lower than those predicted for the 2.7 T discharge due to the increase in synchrotron power by a factor of (5.4/2.7)$^2$ = 4. Additionally, the poloidal magnetic field resulting from the higher plasma current ($I_P$ = 1 MA) is strong enough to confine REs on each surface. An increase in $n_{RE}$ on inner surfaces is predicted at $t \approx$ 0.6 s, about the time that $E/E_C \geq 5$ and $T_e$ increases in the core plasma. The $n_{RE}$ profile still peaks in the core, but the ratio of $n_{RE}$ on-axis to that on the $q$~=~2 surface is only $\sim$100 in this discharge, which means that high energy REs at the edge contribute more significantly to the total brightness. 

\begin{figure}[h!]
	\centering
		\includegraphics[width=\linewidth]{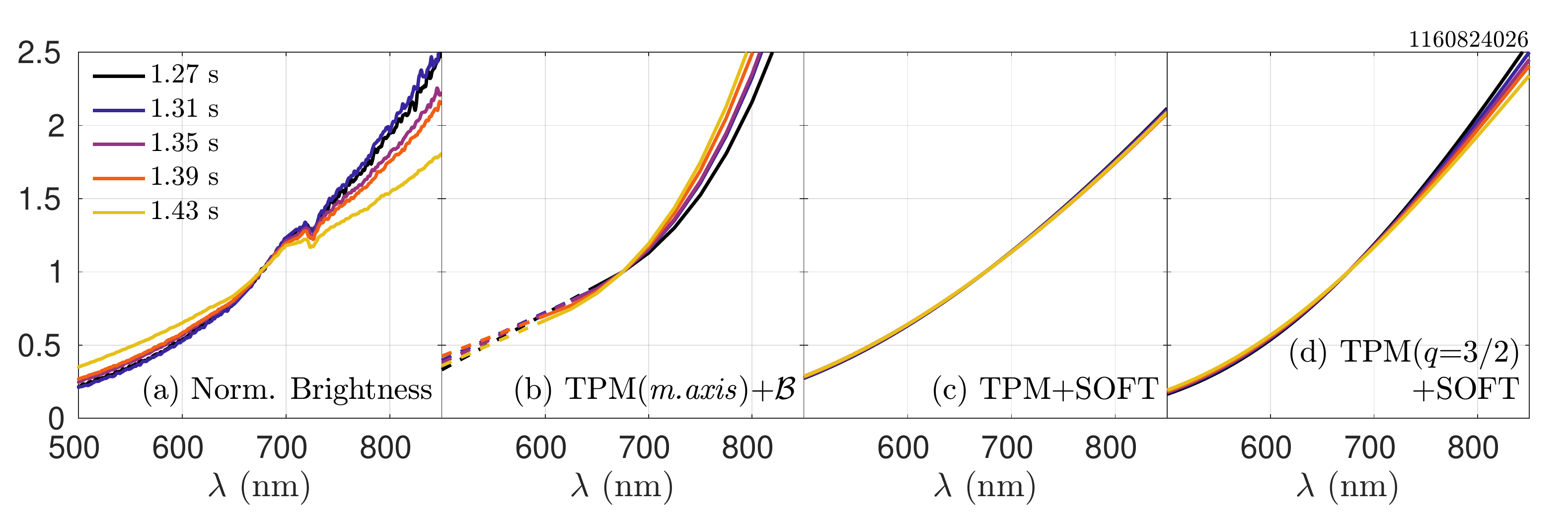}
	\caption{For five times during the 5.4 T discharge: comparisons of (a) experimental normalized brightness spectra to synthetic data from (b) TPM data only on-axis input into $\mathcal{B}$ (\ref{eq:brightness}), (c) TPM data from all surfaces input into SOFT, and (d) TPM data from only the $q$~=~3/2 surface input into SOFT. The dashed lines in (b) are linear extrapolations of the spectral curves to wavelengths at which the oscillating integrand of $\mathcal{P}$ causes large errors.}
	\label{fig:normBrightness54T}
\end{figure}

Synchrotron spectra were measured by the CCW spectrometer for this discharge. The experimental and synthetic normalized brightnesses are compared at five times in figure~\ref{fig:normBrightness54T}. While there is significant variation in measured spectra in subplot (a), almost none is predicted by the TPM results input into SOFT, as shown in subplot (c). This is due to TPM data changing little over the times of interest, which is seen for energy, pitch, and density in figure~\ref{fig:plasmaParams54T}f-h. Nevertheless, at this magnetic field, the brightness formula $\mathcal{B}$ (figure~\ref{fig:normBrightness54T}b) is no longer able to reproduce experimental spectra. Note that the TPM+$\mathcal{B}$ model predicts far more emission at wavelengths $\lambda >$ 675 nm compared to $\lambda <$ 675 nm, and the time evolution is opposite to that measured. The SOFT contribution from only the $q$~=~3/2 surface is shown in figure~\ref{fig:normBrightness54T}d to highlight that this data better matches earlier times, while that in subplot (b) is more similar to later times. A more precise prediction of the spatial density profile could be needed to best match experiment. Even so, SOFT better predicts the spectral shape and evolution than the traditional approach.

\subsection{$B_0$ = 7.8 T}
 
\begin{figure}[h!]
	\centering
		\includegraphics[width=0.75\linewidth]{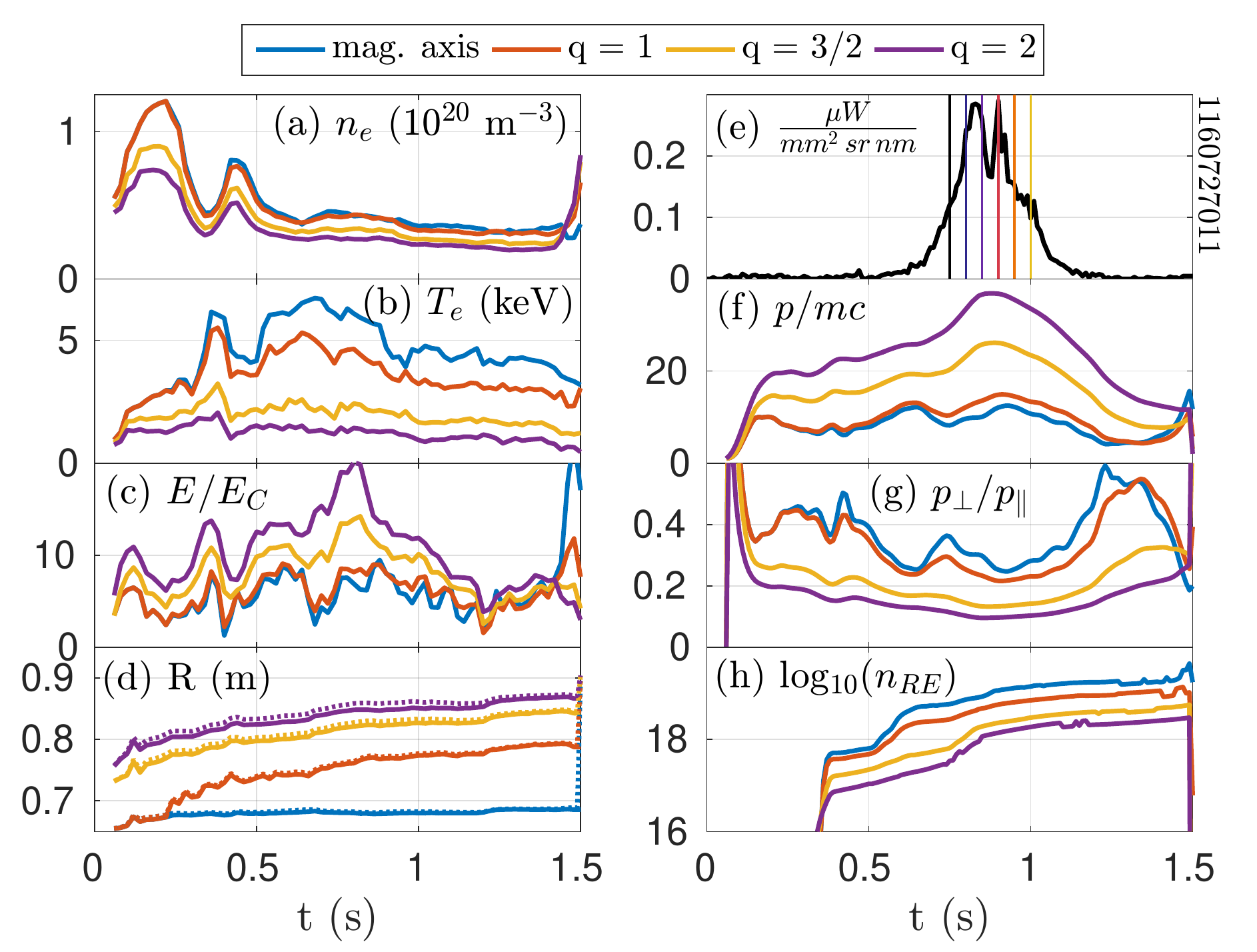}
	\caption{For the 7.8 T discharge, parameters on the magnetic axis and $q$~=~1, 3/2, and 2 surfaces: (a) electron densities, (b) electron temperatures, (c) ratios of electric to critical fields, and (d) outer midplane major radii with (dotted) and without (solid) including drifts; (e) measured synchrotron brightness (at $\lambda$ = 850 nm) with times of interest corresponding to figure~\ref{fig:normBrightness78T} indicated as vertical lines; and TPM results for RE (f) total normalized momenta, (g) pitches, and (h) densities. ($Z_{\mathrm{eff}} = 4$)}
	\label{fig:plasmaParams78T}
\end{figure}

While the same procedure of decreasing $n_e$ was employed in the 7.8 T discharge, synchrotron radiation was only observed after the ramp in $I_P$ = 1.0-1.4 MA from $t$~=~0.6-0.8 s. The resulting increase in $E/E_C$ is most significant on the outer surfaces, $q$~=~3/2 and 2 (figure~\ref{fig:plasmaParams78T}c). The TPM-predicted energies peak at $t \sim$ 0.8 s and then decay, which follow quite closely the evolution of synchrotron brightness (figure~\ref{fig:plasmaParams78T}e). However, the sharp ``dip" in experimental brightness around $t \approx$ 0.88 s is not captured in the TPM. Instead, the start of a locked mode at this time likely causes increased transport of REs which would otherwise be confined by the relatively high plasma current. Nonetheless, predicted synchrotron spectra after this time match experiment quite well, suggesting that the large amplitude MHD activity affected RE density more than energy distribution in this case. The predicted RE densities are also much closer in magnitude than for the other discharges, differing by only a factor of $\sim$10 between the magnetic axis and $q$~=~2 surface. Thus, for this discharge, we expect an even greater contribution to the total synchrotron emission from high energy REs on outer flux surfaces. Note also that $n_{RE}$ is predicted to continue increasing even after the decline in observed synchrotron emission at $t \approx$ 1.1 s. This is consistent with HXR data (figure~\ref{fig:plasmaParameters}f-g) up to the time of disruption; however, RE energies are simply not large enough to produce significant visible synchrotron radiation during these later times.
 
\begin{figure}[h!]
	\centering
		\includegraphics[width=\linewidth]{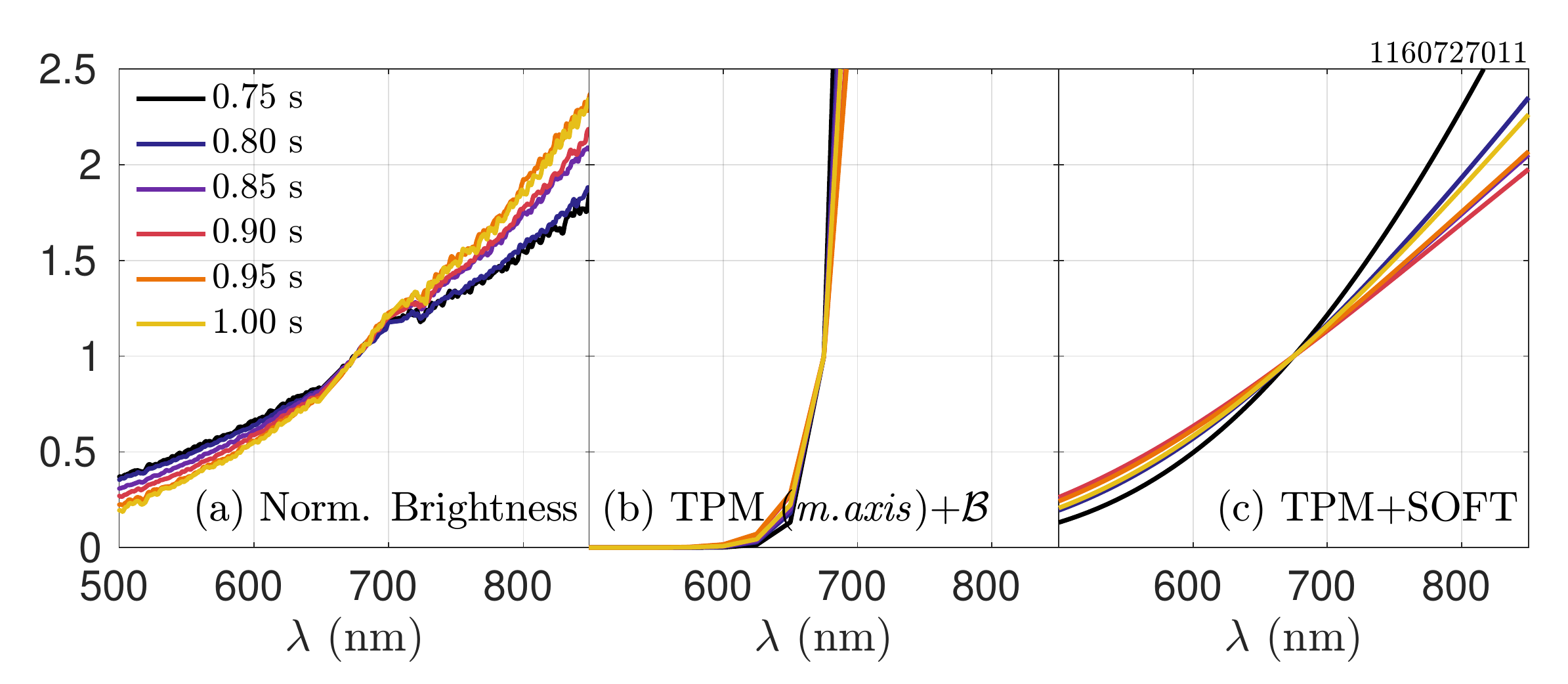}
	\caption{For six times during the 7.8 T discharge: comparisons of (a) experimental normalized brightness spectra to synthetic data from (b) on-axis TPM data input into $\mathcal{B}$ (\ref{eq:brightness}), and (c) TPM data from all surfaces input into SOFT. Note that in (b), equation~(21) in \cite{pankratov1999} was used as a better approximation for $\mathcal{P}(\lambda,\vec{p},t)$.}
	\label{fig:normBrightness78T}
\end{figure}

The 7.8 T discharge was run in reversed-field configuration, i.e. $I_P$ and $B$ opposite to those in the 2.7 and 5.4 T discharges; therefore, synchrotron emission was measured by the CW spectrometer The measured normalized brightnesses at six times are shown in figure~\ref{fig:normBrightness78T}a. Note that the ``slope" \textit{increases} in time, opposite to the trends of the 2.7 and 5.4 T discharges. Using TPM data on the magnetic axis and brightness formula $\mathcal{B}$ (\ref{eq:brightness}) produces spectra completely unlike experiment, as seen in figure~\ref{fig:normBrightness78T}b. One explanation for this is that an asymptotic expansion of $\mathcal{P}(\lambda,\vec{p},t)$, given by equation (21) in \cite{pankratov1999}, is used to compute the spectra, as the full calculation becomes increasingly computationally-intensive for such low energies. More importantly, as seen in figure~\ref{fig:cmod}, synchrotron emission from REs on the magnetic axis is never directly viewed by the spectrometer, so emission from REs on other flux surfaces \textit{must} be considered. (SOFT also determines zero contribution to the spectra from REs on-axis.) Coupling the TPM with SOFT (figure~\ref{fig:normBrightness78T}c) is better at matching experiment, but only for later times $t$ = 0.85-1.00 s. Earlier times $t$ = 0.75-0.80 s are opposite the experimental trend, and the time evolution of the ``slope" is not monotonic, unlike the measured data. Because all REs should be well-confined by the high $I_P$ = 1.4 MA, this cannot be easily explained. Implementation of full drift orbit effects into SOFT is likely needed to best reproduce experimental spectra.

\section{Discussion}\label{sec:discussion}

\begin{figure*}[h!]
	\centering
	\includegraphics[width=\linewidth]{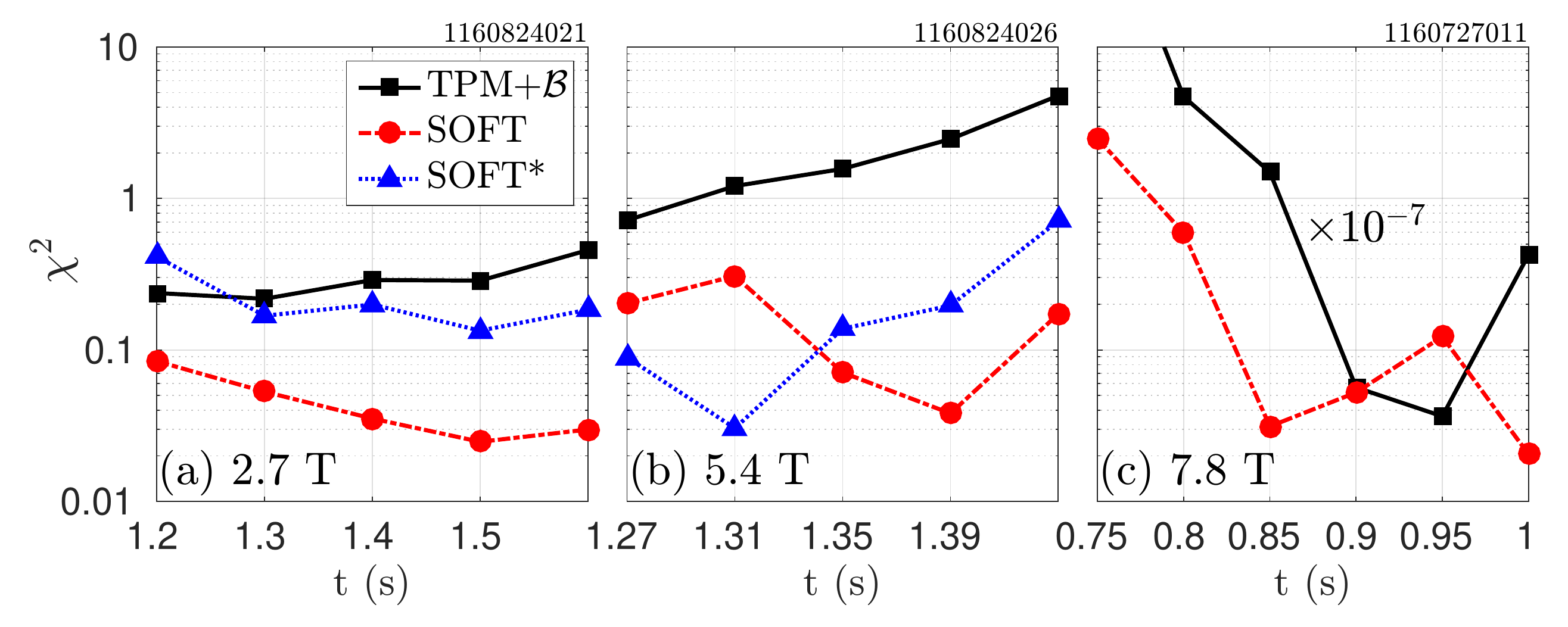}
	\caption{Time evolution of $\chi^2$ values comparing experimental to synthetic spectra for the (a) 2.7, (b) 5.4, and (c) 7.8 T discharges, for which the SOFT (dot-dashed, circles) and TPM+$\mathcal{B}$ (solid, squares) correspond to figures~\ref{fig:normBrightness27T}, \ref{fig:normBrightness54T}, and \ref{fig:normBrightness78T} subplots (b) and (c), respectively. The SOFT* data (dotted, triangles) highlight additional SOFT simulations for comparison: Here, subplot (a) uses CODE distributions input into the SOFT cone model from figure~\ref{fig:normBrightness27T}d, and subplot (b) includes the $q$~=~3/2 surface TPM contribution from figure~\ref{fig:normBrightness54T}d. Note in (a) that $n_{RE}$($t$ = 1.23 s) was used for the SOFT data, and in (c) that the TPM+$\mathcal{B}$ curve has been scaled by the factor given.}
	\label{fig:X2}
\end{figure*}

The $\chi^2$ values comparing synthetic and experimental normalized brightnesses are shown in figure~\ref{fig:X2} for the three discharges. Assuming synchrotron emission is dominated by REs on-axis and using a simplified brightness formula (i.e. TPM+$\mathcal{B}$) works well for the 2.7 T discharge, but becomes increasingly inaccurate for the higher field discharges. The agreement in the low $B$ case is likely due to REs at outer surfaces losing confinement (due to low $I_P$) so that emission was in fact dominated by core REs. As higher magnetic fields lead to increased synchrotron radiation, lower energies are attained by REs and make the calculation of $\mathcal{P}$ and thus $\mathcal{B}$ less precise. A synthetic diagnostic like SOFT -- which incorporates RE momenta and spatial profiles, magnetic equilibria, and detector geometry and spectral range -- is required to reproduce experimental spectra. 

\begin{figure}
	\centering
		\includegraphics[width=0.5\linewidth]{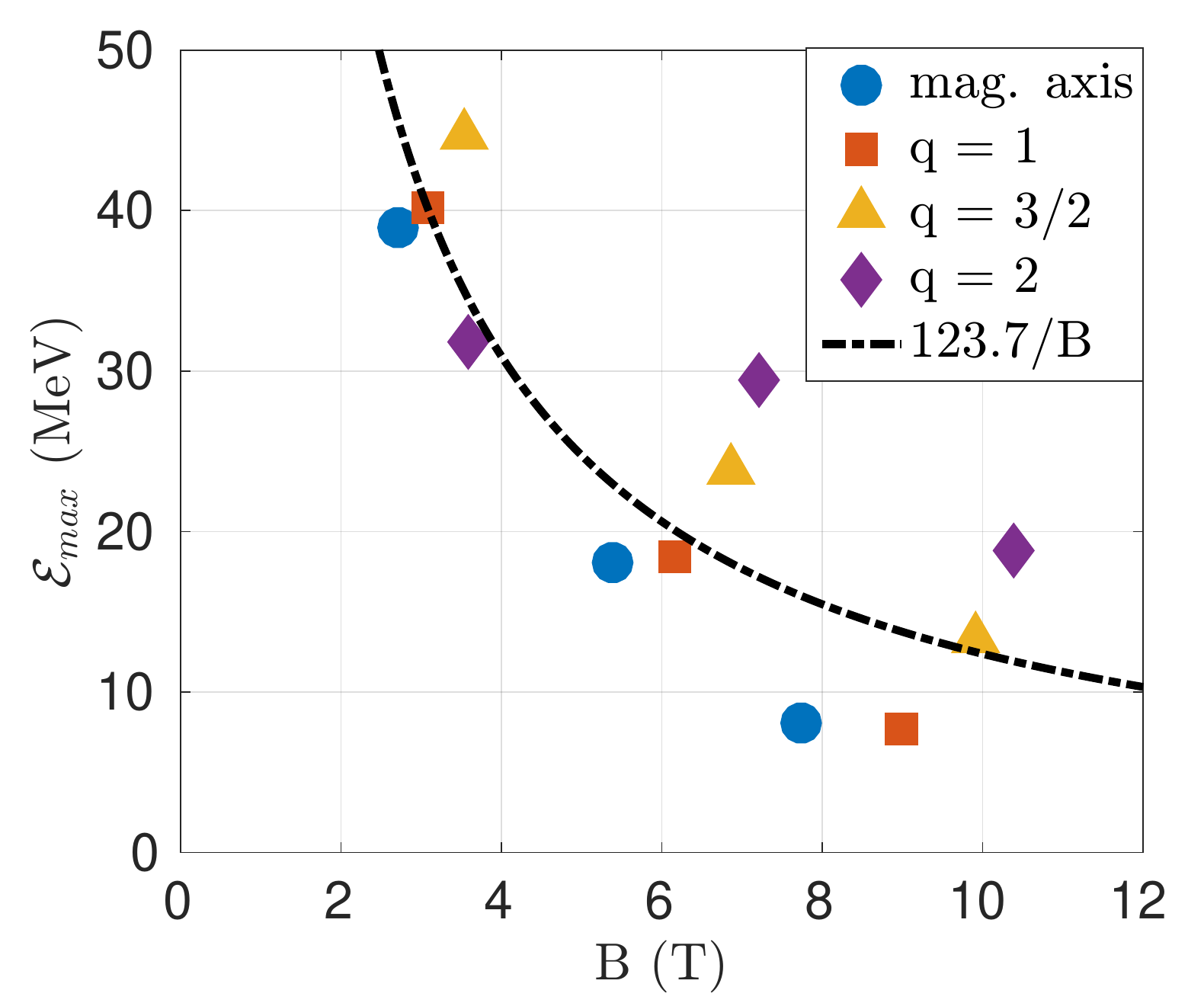}
	\caption{Maximum RE energy predicted by the TPM plotted as a function of highest magnetic field experienced for particles on the magnetic axis (circles) and $q$~=~1 (squares), 3/2 (triangles), and 2 (diamonds) surfaces for the 2.7, 5.4, and 7.8 T discharges. A representative best-fit curve of the form $\mathcal{E}_{max} \propto 1/B$ is shown.}
	\label{fig:maxEnergy}
\end{figure}

In addition, the reproducibility of experimental spectra validates the methodology used which incorporated both TPM energy and density evolutions as well as drift orbit effects on several flux surfaces throughout the plasma. The maximum energies attained by REs during flattop $I_P$, as calculated by the TPM, are shown in figure~\ref{fig:maxEnergy} as a function of highest magnetic field experienced by the particle. Using a simple assumption of constant radiated synchrotron power, i.e. $p_\perp^2 B^2 \approx$ constant, a function of the form $\mathcal{E}_{max} \sim 1/B$ is fit to all data. In C-Mod discharges with similar $E/E_C$ values, experimental data are consistent with lower RE energies attained at higher magnetic fields, as synchrotron radiation becomes a more important power loss mechanism. A similar result was inferred from HXR emission from REs in DIII-D \cite{paz-soldan2017}.

Moreover, note from figure~\ref{fig:maxEnergy} that at higher magnetic fields, REs on outer flux surfaces attained higher energies than those near the plasma core. This results from a combination of higher electric fields (due to $E \sim 1/R$ and diffusion time into the plasma), lower plasma densities at the edge, and increased confinement from larger $I_P$ (which is allowed by higher $B$ at a fixed edge safety factor, $q_{95}$). While a greater number of REs could be generated near the magnetic axis due to higher temperatures, a full study of RE energy and density evolution throughout the plasma is necessary for the prediction of RE dynamics and the threat they pose to future tokamaks.

\section{Summary}\label{sec:summary}

Additional RE experiments were performed at high magnetic fields on Alcator C-Mod. The threshold electric field for RE generation was deduced from HXR and photo-neutron measurements and compared to recent theoretical predictions. This effective critical field was found to be a factor of $\sim$5 higher than that predicted by purely collisional theory \cite{connor1975} and $\sim$3 times higher than estimates requiring knowledge of only bulk plasma parameters, i.e. electric field, electron density, effective charge, and magnetic field \cite{martin-solis2010,aleynikov2015}. Knowledge of the approximate RE energy and pitch allows a better prediction of the threshold field \cite{stahl2015}, but may not be possible for future fusion devices. Therefore, further work should be done to predict the threshold electric field for RE onset, particularly for high magnetic field scenarios.

Absolutely-calibrated visible spectrometers were also installed on C-Mod to measure RE synchrotron radiation spectra. In particular, the spectra from three plasma discharges with similar electric-to-friction force ratios ($E_0/E_C$) but varying magnetic fields ($B_0$ = 2.7, 5.4, and 7.8 T) were studied. A test particle model (TPM) of RE energy \cite{martin-solis1998} and density \cite{connor1975,rosenbluth1997} evolutions was used to estimate RE dynamics on the magnetic axis and $q$~=~1, 3/2, and 2 flux surfaces. Drift orbits and loss of confinement were also included \cite{knoepfel1979}. A synthetic diagnostic SOFT \cite{hoppe2017} was employed to produce synthetic spectra including contributions throughout the plasma for given magnetic geometries and spectrometer specifications. It was found that experimental spectra could be best reproduced when incorporating the spatiotemporal results of the TPM into SOFT, especially at high magnetic fields. Simple calculations of the expected spectra, as in (\ref{eq:brightness}), will lead to unphysical interpretations of RE energies. In this study, the experimental spectra were consistent with RE energies decreasing when toroidal magnetic field was increased, due to more power lost by synchrotron radiation. This motivates further exploration of high $B$-field fusion devices for which high energy REs could be of lesser concern.

\section*{Acknowledgements}

The authors thank A. Stahl for fruitful discussions and the Alcator C-Mod team. This work was supported by US DOE Grant DE-FC02-99ER54512, using Alcator C-Mod, a DOE Office of Science User Facility; Vetenskapsr\aa det (Dnr 2014-5510); and the European Research Council (ERC-2014-CoG grant 647121). 

\section*{References}
\bibliographystyle{unsrt}
\bibliography{bib}

\end{document}